\documentclass[aps,prb,twocolumn,superscriptaddress]{revtex4}
\usepackage{graphics,epsfig,amssymb}

\newcommand{\beq}{\begin{equation}}
\newcommand{\eeq}{\end{equation}}
\newcommand{\bea}{\begin{eqnarray}}
\newcommand{\eea}{\end{eqnarray}}
\newcommand{\beas}{\begin{eqnarray*}}
\newcommand{\eeas}{\end{eqnarray*}}

\newcommand{\pdag}{{\phantom{\dagger}}}
\newcommand{\nn}{\nonumber}
\newcommand{\eps}{\epsilon^\pdag}
\newcommand{\ds}{\displaystyle}

\begin{document}


\title{Non--Equilibrium Scaling Analysis of the Kondo Model
with Voltage Bias}

\author{Peter Fritsch}
\author{Stefan Kehrein} 
\affiliation{Physics Department, Arnold Sommerfeld Center for Theoretical Physics and Center for NanoScience,
Ludwig-Maximilians-Universit\"at M\"unchen, Theresisenstr.~37, 80333 M\"unchen, Germany} 


\date{\today}

\begin{abstract}
The quintessential description of Kondo physics in equilibrium is obtained within a
scaling picture that shows the buildup of Kondo screening at low temperature. 
For the non-equilibrium Kondo model with a voltage bias the key new feature are
decoherence effects due to the current across the impurity. In the present paper
we show how one can develop a consistent framework for studying the 
non-equilibrium Kondo model within a scaling picture of infinitesimal unitary
transformations (flow equations). Decoherence effects appear naturally in third
order of the $\beta$-function and dominate the Hamiltonian flow for sufficiently
large voltage bias. We work out the spin dynamics in non-equilibrium and compare
it with finite temperature equilibrium results. In particular, we report on the 
behavior of the static spin susceptibility including leading logarithmic corrections
and compare it with the celebrated equilibrium result as a function of temperature. 

\end{abstract}


\maketitle




\section{Introduction}
\subsection{Motivation}
Scaling concepts provide an invaluable tool in modern Theoretical Physics
and are fundamentally important for understanding universal behavior and
phase diagrams of quantum many--body systems. To date, most of these
applications have been to equilibrium problems, where the basic idea of
scaling to focus on the low--energy excitations of the system is most 
naturally applicable. Non--equilibrium problems, like systems prepared in
an initial non--equilibrium state, or systems in a steady state far
from equilibrium (like transport processes between reservoirs), are
already conceptually problematic for a scaling analysis since the focus
cannot be on the low--energy excitations from the equilibrium ground
state alone. A lot of theoretical work has recently been 
devoted to developing extension of scaling ideas to non--equilibrium problems,
namely the frequency--dependent renormalization group \cite{Rosch03,Rosch05},
the real time renormalization group \cite{Schoeller00,Schoeller07},
the Coulomb gas representation \cite{Mitra,Millis}
and the flow equation approach \cite{Kehrein04,HacklKehrein}.

This paper provides a detailed account of the flow equation approach applied
to a non--equilibrium steady state of a quantum impurity model. Concretely,
we study the Kondo model with a dc--voltage bias that produces a stationary
current. This non--equilibrium
Kondo model is a natural candidate for our analysis since it i)~can be realized
experimentally in quantum dot experiments and ii)~its equilibrium version
is the paradigm model for strong--coupling impurity physics in condensed matter
theory. The current paper is a substantial extension of work previously presented in 
Ref.~\cite{Kehrein04} In particular, we study in detail the spin dynamics and
the behavior of the static spin susceptibility~$\chi_{0}$ in non-equilibrium. We extend
previous results in the literature, which, for example, allows us to compare the leading
logarithmic corrections in non-equilibrium with the equilibrium result for~$\chi_{0}(T)$.

The experimental motivation for our study is the observation of the Kondo
effect in the Coulomb blockade regime of quantum dots, first realized
in 1998~\cite{GoldhaberGordon98,Cronenwett98,Schmid98}. If a quantum dot
that is weakly coupled to two leads is tuned into the Coulomb blockade 
regime such that it carries a net spin, resonant tunneling through the
dot leads to a Kondoesque increase of the conductance up to the unitarity
limit upon lowering temperature \cite{Wiel00}. This was first predicted
theoretically in Refs.~\cite{Glazman88,Ng88}. The case of small
voltage bias between the two leads, $V\ll T_{\rm K}$, where $T_{\rm K}$
is the equilibrium Kondo temperature, can be analyzed using linear--response
theory from the well--understood equilibrium ground state \cite{Glazman88,Ng88}. 

The case of intermediate voltage bias, $V\approx T_{\rm K}$, that
matches the linear--response to the large voltage bias regime, has
until recently been out of reach for any controlled theoretical investigation.
Very recently, new methods like
the scattering state numerical renormalization group \cite{Anders08},
the time--dependent density renormalization group \cite{Schmitteckert},
and the scattering state Bethe ansatz \cite{Andrei} have been developed
that can access this crossover regime, though much more work needs
to be done before a complete picture will emerge.

In this paper we study the situation of large voltage bias, 
$V\gg T_{\rm K}$, where one expects to find weak--coupling physics and
therefore the possibility to do a controlled renormalized perturbation expansion.
Kaminski {\it et al.} first developed a scaling picture of the
large voltage bias Kondo model based on the invariance of the 
current under the RG--flow~\cite{Glazman99}. Subsequently, Rosch {\it et al.}
developed a more sophisticated approach based on frequency--dependent
vertices and Keldysh diagrammatics~\cite{Rosch03,Rosch05,Rosch04a,Rosch04b}. In both approaches
decoherence effects due to non--equilibrium spin 
relaxation processes generated by the stationary current play a key role:
The resulting decoherence rate 
$\Gamma_{\rm rel}\propto V/\ln^2(V/T_{\rm K})$,\cite{Rosch01} is essential for cutting
off inter--lead scattering processes that are not immediately cut off
by the voltage bias. The decoherence rate $\Gamma_{\rm rel}$ is therefore 
responsible for actually making the situation $V\gg T_{\rm K}$ a 
weak--coupling problem. Such decoherence effects related to the noise
produced in stationary non--equilibrium states are expected to be 
generically important in non--equilibrium problems. 
 
The flow equation approach applied to the large voltage bias regime
permits an analysis of this regime that is complimentary 
to the other approaches mentioned above: Here
the important decoherence effects emerge in a Hamiltonian scaling 
framework. We will improve the accuracy of some 
quantities using this approach, for example regarding the non--equilibrium dynamical spin
susceptibility and the behavior of the
static spin susceptibility as a function of the voltage bias. 
In addition, the Hamiltonian scaling picture developed here offers insights into non--equilibrium
scaling in general.
 
The basic idea
of the flow equation approach~\cite{Wegner94,Wilson93,Kehreinbook}
is to make a Hamiltonian increasingly more
band--diagonal by a suitable sequence of infinitesimal unitary
transformations. Band--diagonality is here measured in terms of
the energy transfer of scattering processes, therefore in the
flow equation framework the flowing Hamiltonian $H(\Lambda_{\rm feq})$
at the ``scale" $\Lambda_{\rm feq}$ only contains interaction
matrix elements with energy transfer $|\Delta E|\lesssim \Lambda_{\rm feq}$.
This should be compared with the conventional scaling approach
where one integrates out Hilbert space states with an energy 
larger than some cutoff~$\Lambda_{\rm RG}$, and then successively 
lowers this cutoff. The essential idea behind both approaches is
to organize a perturbative expansion in such a way as to first deal
with large energy denominators and to avoid
small energy denominators; this sequence allows for stable expansions
even when dealing with nonperturbative energy scales like the Kondo
temperature~$T_{\rm K}$. 

\begin{figure}[t]
\includegraphics[clip=true,width=8.0cm]{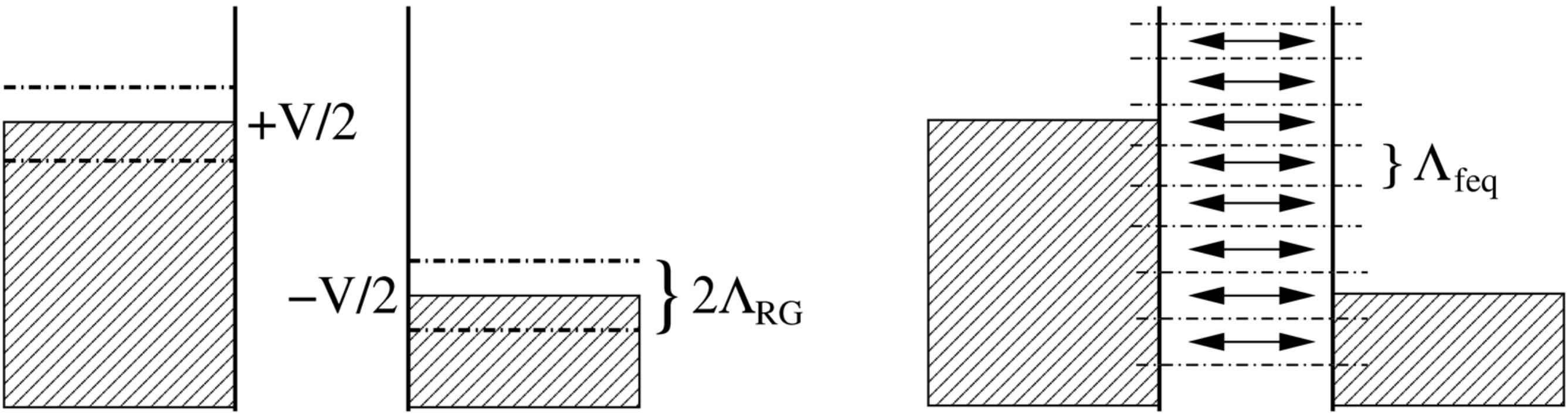}
\caption{Left: Conventional scaling picture where states are integrated
out around the two Fermi surfaces with voltage bias~$V$ 
(here depicted for cutoff~$\Lambda_{\rm RG}<V$).
Right: Flow equation approach. Here all scattering processes with energy
transfer $|\Delta E|\lesssim \Lambda_{\rm feq}$ are retained in 
$H(\Lambda_{\rm feq})$.}
\label{Fig_FEQvsRG}    
\end{figure}

However, one important conceptual difference between these two approaches
is that the flow equation approach retains all states in the Hilbert space
but decouples them,
while the conventional scaling approach actually eliminates high--energy states.  
If one only focuses on the low--energy physics in an equilibrium problem,
this difference is unimportant and the flow equation approach is consistent
with the conventional RG--flow with the identification 
$\Lambda_{\rm feq}\propto\Lambda_{\rm RG}$ (explicit examples for this
observation can be found in Ref.~\cite{Kehrein_SG,Kehrein_KM}). In a non--equilibrium
situation like the Kondo model with voltage bias the difference is, however,
more fundamental. Fig.~\ref{Fig_FEQvsRG} schematically depicts the two schemes once
the cutoff is reduced below the scale set by the voltage bias, $\Lambda<V$.
In the flow equation framework the Hamiltonian $H(\Lambda_{\rm feq})$
still describes the stationary current flowing across the dot, while scaling around
the separate Fermi surfaces of the two leads does not allow for energy--diagonal
transport processes across the dot anymore. Therefore a naive application of the conventional 
scaling approach purports to eliminate scattering processes with small
energy--denominators $|\Delta E|\ll \Lambda_{\rm RG}$ across the dot, 
which violates the spirit of renormalized perturbation theory.
We will later see that it is precisely this point which allows us to understand 
qualitatively and quantitatively non--equilibrium decoherence in a Hamiltonian framework.

\subsection{Outline}

This paper is organized as follows. In Sect.~II we define the model
and briefly review its conventional scaling analysis. In Sect.~III.A
we first introduce the flow equation method, and then in part~III.B
apply it to a general Kondo model Hamiltonian with an arbitrary number
of leads with or without voltage bias. This is the core part of our
calculation. The resulting equations are then analyzed in equilibrium
in part~III.C, where we reproduce the conventional two--loop $\beta$--function.
In part~III.D and III.E we then specialize to the case of interest
in this paper; the Kondo model with two leads and voltage bias~$V$.
Using these flow equations, we work out and analyze the scaling picture of the
Kondo model with voltage bias in Sect.~IV. Sect.~V discusses our
results for the spin dynamics in non--equilibrium, which are compared with
the well--known equilibrium behavior at nonzero temperature. 
In particular Sects.~V.C and V.D contain the key new results of this
work regarding the spin dynamics and the static spin susceptibility 
in non--equilibrium. A summary of results and conclusions is presented in 
Sect.~VI.
Appendix~A contains important commutators
and normal--ordering results that are frequently needed, and 
Appendix~B contains details of the numerical solution of the flow equations.

\section{The Kondo Model with Voltage Bias}
\subsection{Model}
\label{ch_Kondomodel}
We are investigating a quantum dot that can be modelled by a
spin--1/2 degree of freedom~$\vec S$ 
coupled to conduction electrons in a left ($l$) and a right ($r$) lead 
with no external magnetic field
\beq
H=\sum_{a,p,\alpha} (\epsilon^\pdag_p-\mu^\pdag_a)
c^\dag_{ap\alpha} c^\pdag_{ap\alpha}\!
+\sum_{a',a} J^\pdag_{a'a} \sum_{p',p} \vec S\, \cdot\, \vec s_{(a'p')(ap)} \ .
\label{Kondo_nonequ}
\eeq
Here $a',a=l,r$ label the two leads, $p',p$ are momentum labels,
and the chemical potentials are given by $\mu_{l,r}=\pm V/2$.
The conduction band electron spin operators are defined by
$\vec s_{(a'p')(ap)}=\frac{1}{2}\sum_{\alpha,\beta} c^\dag_{a'p'\alpha}
\vec\sigma^\pdag_{\alpha\beta} c^\pdag_{ap\beta}$ where $\vec\sigma$
are the Pauli matrices. The couplings $J_{a'a}$ describe the
exchange interaction with the localized spin degree of freedom,
with $J_{lr}=J_{rl}$ for hermiticiy.
If the quantum dot can be described by an Anderson impurity model
with tunneling rates $\Gamma_{l,r}$ from the left/right lead, the coupling
constants of our effective Kondo model are related by 
$J_{lr}^2=J^\pdag_{ll} J^\pdag_{rr}$
and $J_{ll}/J_{rr}=\Gamma_l/\Gamma_r$ \cite{Glazman99}. 
We define $r\stackrel{\rm def}{=}\Gamma_l/\Gamma_r$ as the {\em asymmetry
parameter} of our model. We
will derive the scaling equations without making any of the above assumptions
so that also more general complex quantum dot structures can 
be investigated (e.g., in double dot systems the relation
$J_{lr}^2=J^\pdag_{ll} J^\pdag_{rr}$ is not necessarily fulfilled). We assume
a linear dispersion relation leading to 
a constant conduction band density of states~$\rho$ and introduce the
dimensionless coupling constants $g_l=\rho J_{ll}, g_r=J_{rr}$
and $g_t=\rho J_{lr}$ ($t$ for transport). The band cutoff is denoted
by~$D$ and we are only interested in the universal behavior on energy
scales much smaller than~$D$.

\subsection{Conventional Scaling Analysis}
Below we will sum up the results of the conventional scaling analysis
for the non--equilibrium Kondo model in the spirit of Fig.~\ref{Fig_FEQvsRG},
that is reminiscent of Anderson's ``poor man's scaling" in equilibrium 
\cite{Anderson70}. The derivation of the one--loop results is straightforward,
see Refs.~\cite{Glazman99,Coleman01}. Upon lowering the
cutoff~$\Lambda_{\rm RG}$ around the two Fermi surfaces, but while 
$\Lambda_{\rm RG}\gtrsim V$ still holds, one finds the equilibrium scaling
equations
\bea 
\frac{dg_l}{d\ln\Lambda_{\rm RG}}&=&-g_l^2-g_t^2 \\
\frac{dg_r}{d\ln\Lambda_{\rm RG}}&=&-g_r^2-g_t^2 \\
\frac{dg_t}{d\ln\Lambda_{\rm RG}}&=&-g_t(g_l+g_r)
\label{eq_Glazman99} 
\eea
plus third order terms in the coupling constants. Once the cutoff is
reduced below the voltage bias, $\Lambda_{\rm RG}\lesssim V$, the
strong--coupling scaling of the coupling $g_t$ stops since
there is no sharp Fermi surface for transport processes on this scale.
Similarly, only the inter--lead scattering processes still contribute
to the scaling equations of $g_l$ and $g_r$
\bea 
\frac{dg_l}{d\ln\Lambda_{\rm RG}}&=&-g_l^2 \label{eq_poorman1} \\
\frac{dg_r}{d\ln\Lambda_{\rm RG}}&=&-g_r^2 \label{eq_poorman2} \\
\frac{dg_t}{d\ln\Lambda_{\rm RG}}&=&0 \ . 
\eea
We now assume that the Kondo model can be derived from an underlying
Anderson impurity model, and one easily solves the equation for $g_t$ 
down to the infrared limit~\cite{Glazman99}
\beq
g_t(\Lambda_{\rm RG}=0)=\frac{\sqrt{\Gamma_l \Gamma_r}}{\Gamma_l+\Gamma_r}\:
\frac{1}{\ln(V/T_{\rm K})} \ ,
\label{eq_gtIR}
\eeq
where $T_{\rm K}=D\,\exp(-1/g_l+g_r)$ is the Kondo temperature of the
equilibrium model (i.e., for $V=0$). The IR--coupling (\ref{eq_gtIR}) 
determines the current~$I$ across the dot and a conventional second order
Keldysh calculation in the renormalized quantities yields~\cite{Glazman99}
$I=(3\pi/4)\,V\,g_t^2(\Lambda_{\rm RG}=0)$. This leads to perturbative
result for the differential conductance~$G(V)$ valid in the limit $V\gg T_{\rm K}$
\beq
G(V)=G_u\:\frac{3\pi^2}{16\,\ln^2(V/T_{\rm K})} \ ,
\label{eq_GV}
\eeq
where 
\bea
G_u&=&\frac{2e^2}{h}\: \frac{4\Gamma_l \Gamma_r}{(\Gamma_l+\Gamma_r)^2}  \nn \\
&=& \frac{2e^2}{h}\: \frac{4}{(1+r)(1+r^{-1})} 
\label{eq_Gu}
\eea
is the conductance in the unitarity limit.

Notice that there is no IR--cutoff mechanism for the inter--lead scattering
processes in the scaling equations (\ref{eq_poorman1}--\ref{eq_poorman2}) 
for $g_l, g_r$, which therefore yield strong--coupling
divergences even for large voltage bias~\cite{Coleman01}. However, Rosch {\it et al.} have
pointed out the importance of spin relaxation processes due to the
stationary current.\cite{Rosch01} Second order perturbation theory in the renormalized
quantities yields the decoherence rate
\beq
\Gamma_{\rm rel}\propto \frac{V}{\ln^2(V/T_{\rm K})} \ ,
\eeq
which cuts off the inter--lead strong--coupling flow and eliminates
the possibility of two--channel Kondo physics in Kondo dots that can be
derived from an underlying Anderson impurity model.
The observation that the non--equilibrium Kondo model becomes a weak--coupling
problem for $\Gamma_{\rm rel}\gg T_{\rm K}$ has then been exploited to derive
a number of physically relevant quantities like the conduction electron $T$-matrix, the non--equilibrium
magnetization and the conductance both with and without an external magnetic 
field.\cite{Rosch03,Rosch05,Rosch04a,Rosch04b}

\section{Flow Equation Analysis}
\subsection{Flow Equation Method}
The basic idea of the flow equation approach~\cite{Wegner94,Wilson93,Kehreinbook} is
to make a many--particle Hamiltonian increasingly more diagonal through
a sequence of infinitesimal unitary transformations. Such a flow can
be generated by the differential equation
\beq
\frac{dH(B)}{dB}= [\eta(B),H(B)]
\label{eq_feq}
\eeq
with some suitable antihermitean generator~$\eta(B)$; Eq.~(\ref{eq_feq})
then generates a one--parameter family of unitarily equivalent 
Hamiltonians~$H(B)$. We set $H(B=0)$ as the initial Hamiltonian and want
$H(B=\infty)$ to be the final diagonal Hamiltonian. In order to generate
a stable expansion it is of fundamental importance to properly
deal with energy--scale separation during the flow, similar to 
conventional scaling approaches. For small~$B$ (initial phase of the 
flow) we will decouple modes with large energy differences, while
later for large flow parameter~$B$ we will deal with increasingly 
more energy--diagonal processes. This sequence of transformations
is generically generated by the ``canonical" generator suggested
by Wegner~\cite{Wegner94}
\beq
\eta(B)\stackrel{\rm def}{=} [H_0(B),H_{\rm int}(B)] \ ,
\label{eq_eta}
\eeq
where $H_0(B)$ is the diagonal part of the Hamiltonian~$H(B)$ and
$H_{\rm int}(B)$ its interaction part. The flow parameter~$B$ then
has the dimension (Energy)${}^{-2}$, and 
$\Lambda_{\rm feq}\stackrel{\rm def}{=}B^{-1/2}$ is the flow equation
energy scale that expresses how energy--diagonal the Hamiltonian
$H(B)$ has become. Interaction matrix elements with energy transfer
$|\Delta E|\gtrsim \Lambda_{\rm feq}$ are eliminated in $H(\Lambda_{\rm feq})$,
while processes with $|\Delta E|\lesssim \Lambda_{\rm feq}$ are still
retained. We will use the notation with~$B$ or $\Lambda_{\rm feq}$ 
interchangeably.

The flow equation approach has been successfully applied to numerous
equilibrium many--body problems, like dissipative quantum 
systems~\cite{Kehrein96,Kleff}, the two--dimensional Hubbard
model~\cite{Wegner_Hubbard}, low--dimensional spin systems~\cite{Uhrig},
and strong--coupling models like the sine--Gordon model~\cite{Kehrein_SG}
or the Kondo model~\cite{Kehrein_KM,Garst}. Due to its intrinsic 
energy scale separation the flow equation method generates the same
IR--scaling flow as conventional renormalization group methods, with
the identification $\Lambda_{\rm feq}\propto \Lambda_{\rm RG}$. 
Notice that in certain strong--coupling problems~\cite{Kehrein_SG,Kehrein_KM}
the flow equation approach even allows a controlled systematic 
expansion in the strong--coupling phase where
conventional scaling leads to a strong--coupling divergence.

\begin{figure}[t]
\includegraphics[clip=true,width=8.0cm]{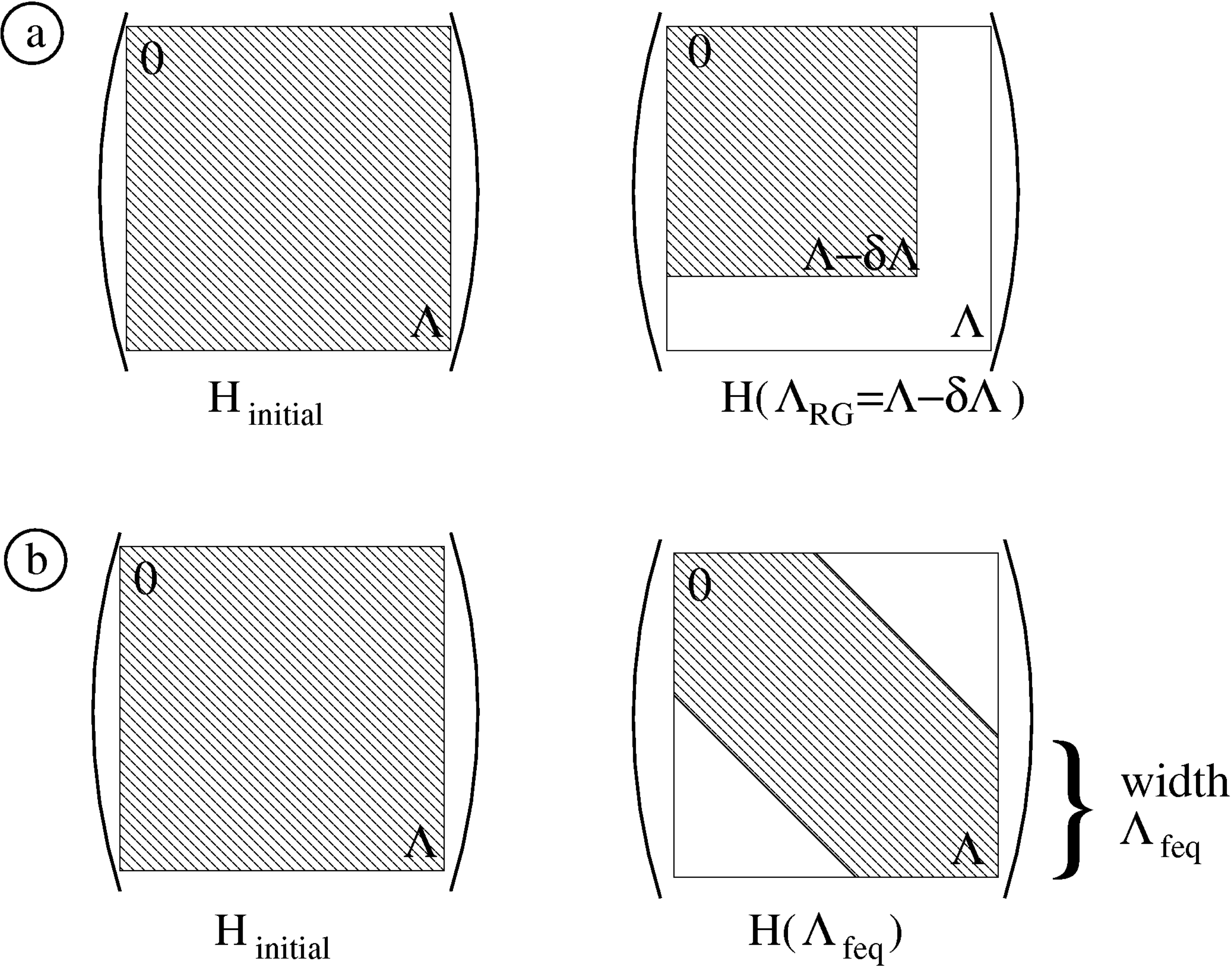}
\caption{a) Schematic picture of the conventional scaling approach 
where states with energies
$[\Lambda_{\rm RG}-\delta\Lambda_{\rm RG},\Lambda_{\rm RG}]$ are 
integrated out. 
b) Flow equation approach. Here all scattering processes with energy
transfer $|\Delta E|\gtrsim \Lambda_{\rm feq}$ are eliminated in 
$H(\Lambda_{\rm feq})$. The shaded areas in a) and b) schematically
indicate nonvanishing interaction matrix elements.}
\label{Fig_FEQvsRG2}    
\end{figure}

The main conceptual difference between flow equations and the conventional
scaling approach is that in the flow equation approach the Hilbert space
remains unchanged while modes become {\em decoupled}, whereas in the
scaling approach high--energy modes are successively integrated out,
compare Fig.~\ref{Fig_FEQvsRG2}.
This allows one to evaluate correlation functions on all energy scales
within the flow equation framework~\cite{Kehrein96,Kehrein_KM}, however, 
for non--equilibrium problems this conceptual difference 
will turn out to be even more fundamental as we will see
later (compare also Fig.~\ref{Fig_FEQvsRG}). Applications of the
flow equation approach to non--equilibrium initial state problems
have been discussed in Refs.~\cite{LobaskinKehrein,HacklKehrein,MoeckelKehrein}. In the sequel, we
will explain in detail the flow equation analysis of
the stationary non--equilibrium problem provided by
the Kondo model with voltage bias. 

\subsection{Kondo Model}
For an interacting many--body problem like the Kondo Hamiltonian
(\ref{Kondo_nonequ}) the flow equation approach generates higher
and higher--order interaction terms in the commutators (\ref{eq_feq})
and (\ref{eq_eta}). Similar to a conventional scaling analysis, 
the flow therefore needs to be approximated by an expansion in a
suitable expansion parameter that remains sufficiently small during
the entire flow. In our problem we will use the running coupling
constants~$J(B)$ as our expansion parameters~\cite{Note_expparameters} 
and perform a consistent systematic expansion in orders of~$J(B)$.
No further input except this systematic expansion is required to
find both coherent {\em and} decoherence effects in our non--equilibrium
problem in a systematic controlled expansion. Since we are performing
unitary transformations (\ref{eq_feq}) which do not explicitly depend
on the ground state of the system (though implicitly as we will see below),
it is actually advantageous to first derive the differential flow
equations for a general Kondo Hamiltonian with couplings to some arbitrary
number of leads (with arbitrary chemical potentials). We introduce
a general index~$t$ labelling the conduction band electrons and write  
\beq
H=\sum_{t,\alpha} \epsilon^\pdag_t\, c^\dag_{t\alpha} c^\pdag_{t\alpha}
+\sum_{t',t} J^\pdag_{t't}\,\vec S\cdot \vec s^\pdag_{t't} 
\label{H_t}
\eeq
with $J_{t't}=J_{tt'}$ for hermiticity.
This will allow us to study both the equilibrium model without voltage bias 
and the non--equilibrium model on the same footing. In the equilibrium model
there is only one lead, therefore $t=p$ with electron momentum~$p$,
whereas in the Kondo model with voltage bias the label $t$ is  
a multi--index labelling both lead left/right and electron momentum,
$t=(a,p)$ with $a=l,r$. 

\subsubsection{$H(B)$ and Generator $\eta(B)$}
During the flow equation procedure new interaction terms are generated
that are not present in the initial Hamiltonian (\ref{H_t}). We will 
neglect the newly generated normal--ordered
terms in $O(J^3)$ and higher in $H(B)$, which can only affect our 
scaling equations for the couplings $J(B)$ in order~$J^4$ and higher (since
the commutator (\ref{eq_feq}) always increases the order by at least one).
We will later see that this order of the flow equation calculation is
the minimum order necessary to obtain a controlled expansion in the sense that the
flowing coupling constants remain finite for $V\gg T_{\rm K}$.
Higher order calculations beyond our present work can be understood as 
corrections to the results presented here.

Within our order of the calculation, the flowing Hamiltonian can be
parametrized as 
\bea
H(B)&=&\sum_{t,\alpha} \epsilon^\pdag_t\,c^\dag_{t\alpha} c^\pdag_{t\alpha}
+\sum_{t',t} J^\pdag_{t't}(B)\, \vec S \cdot \vec s^\pdag_{t't} 
\label{HB} \\
&&+i\sum_{t',t,u',u} K^\pdag_{t't,u'u}(B)\,
:\vec S\cdot (\vec s^\pdag_{t't} \times \vec s^\pdag_{u'u}): \nn \ ,
\eea  
where $K^\pdag_{t't,u'u}(B)$ will turn out to be the only newly generated term
in $O(J^2)$. It obeys $K^\pdag_{t't,u'u}(B)=-K^\pdag_{tt',uu'}(B)$ for
hermiticity and initially $K^\pdag_{t't,u'u}(B=0)=0$. In (\ref{HB})
$:\ldots :$ denotes normal--ordering with respect to the ground state
of the system without Kondo impurity. We use the normal--ordering procedure
as introduced in Ref.~\cite{Wegner94} (for more details see the appendix). 
Notice that we do not need to explicitly normal--order the term
proportional to~$J^\pdag_{t't}$ in (\ref{HB}) since it automatically
appears in its normal--ordered form.
 
The generator of the infinitesimal unitary transformations is given by
(\ref{eq_eta}) where  the diagonal part of $H(B)=H_0+H_{\rm int}(B)$ is
the conduction band kinetic energy
$H_0=\sum_{t,\alpha} \epsilon^\pdag_t\,c^\dag_{t\alpha} c^\pdag_{t\alpha}$.
This leads to
\bea
\eta&=&\sum_{t',t} \eta^{(1)}_{t't}\, \vec S \cdot \vec s^\pdag_{t't} \nn \\
&&+i\sum_{t',t,u',u} \eta^{(2)}_{t't,u'u}\,
:\vec S\cdot (\vec s^\pdag_{t't} \times \vec s^\pdag_{u'u}): 
\eea
where
\bea
\eta^{(1)}_{t't}&=&(\epsilon^\pdag_{t'}-\epsilon^\pdag_t)\,J^\pdag_{t't} \\
\eta^{(2)}_{t't,u'u}&=&
(\epsilon^\pdag_{t'}+\epsilon^\pdag_{u'}-\epsilon^\pdag_t-\epsilon^\pdag_u)\,
K^\pdag_{t't,u'u} \ .
\eea
Here and in the sequel we will usually suppress the explicit $B$--dependence
of all parameters and coupling constants in our notation. 

\subsubsection{Terms in $O(J)$ and $O(J^2)$}
In order to work out the unitary flow of the Hamiltonian (\ref{eq_feq}) 
we now need to calculate the commutator $[\eta(B),H(B)]$. This is the main
calculational problem of the flow equation approach. For the current problem
we will now proceed order by order in the coupling constant~$J$ with this
calculation.

Let us first evaluate $[\eta(B),H_0]$. This is straightforward and leads
to
\bea
[\eta,H_0]&=&\sum_{t',t} (\epsilon^\pdag_{t}-\epsilon^\pdag_{t'})\,
\eta^{(1)}_{t't}\, \vec S \cdot \vec s^\pdag_{t't} 
\label{etaH0} \\
&&+i\sum_{t',t,u',u} 
(\epsilon^\pdag_{t}+\epsilon^\pdag_{u}-\epsilon^\pdag_{t'}-\epsilon^\pdag_{u'})\,
\eta^{(2)}_{t't,u'u} \nn \\
&&\qquad\qquad\times
:\vec S\cdot (\vec s^\pdag_{t't} \times \vec s^\pdag_{u'u}): \ . \nn
\eea
Next we need to find $[\eta(B),H_{\rm int}(B)]$. We first identify its
contributions in order~$J^2$, which can only come from
\beq
C_2=[\sum_{t',t} \eta^{(1)}_{t't}\, \vec S \cdot \vec s^\pdag_{t't}\, , \:
\sum_{u',u} J^\pdag_{u'u}\, \vec S \cdot \vec s^\pdag_{u'u}] \ .
\label{eq_C2}
\eeq
The basic commutator 
$[\vec S \cdot \vec s^\pdag_{t't}\, , \,\vec S \cdot \vec s^\pdag_{u'u}]$
is worked out in the appendix and given in (\ref{app_comm1}).
Inserting it in (\ref{eq_C2}) yields
\bea
C_2&=&i\,\sum_{t',t,u,u'} (\epsilon^\pdag_{t'}-\epsilon^\pdag_t)
J^\pdag_{t't}   J^\pdag_{u'u}
\,:\vec S\cdot (\vec s^\pdag_{t't} \times \vec s^\pdag_{u'u}):
\label{C2} \\
&&+\sum_{t',t,v} (\epsilon^\pdag_{t'}+\epsilon^\pdag_t-2\epsilon^\pdag_v)
J^\pdag_{t'v}   J^\pdag_{vt} \,(n(v)-1/2)\,
\vec S \cdot \vec s^\pdag_{t't} \nn \\
&&+\frac{3}{16}\,\sum_{t',t,v,\alpha} 
(\epsilon^\pdag_{t'}+\epsilon^\pdag_t-2\epsilon^\pdag_v)
J^\pdag_{t'v}   J^\pdag_{vt}
\,:c^\dag_{t'\alpha} c^\pdag_{t\alpha}: \nn 
\eea
plus an uninteresting constant that does not contribute to the further flow.
Here we have defined the expectation value with respect to the
non--interacting ground state
\bea
n(v) &\stackrel{\rm def}{=}& \langle c^\dag_{v\alpha} c^\pdag_{v\alpha}
\rangle \ ,
\eea
which is just the occupation number. No summation over $\alpha$ is
implied in this definition. 
  
Comparing coefficients in (\ref{HB}), (\ref{etaH0}) and (\ref{C2}) we find
the following differential flow equations
\bea
\frac{dJ^\pdag_{t't}}{dB}&=&-(\epsilon^\pdag_{t'}-\epsilon^\pdag_t)^2
J^\pdag_{t't}
\label{eq_J1loop} \\
&&+\sum_v  (\epsilon^\pdag_{t'}+\epsilon^\pdag_t-2\epsilon^\pdag_v)
J^\pdag_{t'v}   J^\pdag_{vt} \,(n(v)-1/2) \nn \\
&&+O(J^3) \nn \\
\frac{dK^\pdag_{t't,u'u}}{dB} &=&
-(\epsilon^\pdag_{t'}+\epsilon^\pdag_{u'}-\epsilon^\pdag_t-\epsilon^\pdag_u)^2\,
K^\pdag_{t't,u'u} \\
&&-(\epsilon^\pdag_{u'}-\epsilon^\pdag_u)
J^\pdag_{t't}   J^\pdag_{u'u} \nn \\
&& +O(J^3) \nn
\eea
From (\ref{C2}) one would conclude that a new term
\beq
\sum_{t',t} \sum_\alpha V^\pdag_{t't}(B)\, :c^\dag_{t'\alpha}
c^\pdag_{t\alpha}:
\label{potscat}
\eeq
is generated in the Hamiltonian $H(B)$ that has not been included so far
\beq
\frac{dV^\pdag_{t't}}{dB} =\frac{3}{16}\, \sum_v  (\epsilon^\pdag_{t'}+\epsilon^\pdag_t-2\epsilon^\pdag_v)
J^\pdag_{t'v}   J^\pdag_{vt} + O(J^3) \ .
\label{feq_V}
\eeq
However, we will see below that this contribution vanishes identically due to 
symmetry reasons (in fact, it only appears in $O(J^4)$).

\subsubsection{Terms in $O(J^3)$}
Our goal is to derive the $\beta$--function of the Kondo model up to two--loop
order within the flow equation framework. This implies that we need to
calculate the terms in order~$J^3$ that contribute to the differential
equation (\ref{eq_J1loop}) for $J^\pdag_{t't}(B)$. Notice that our differential
equation for $K^\pdag_{t't,u'u}(B)$ generates this term (which is not
present for $B=0$) in order~$J^2$. Also notice that
we do not need to calculate its $O(J^3)$--contribution because this can only
feed back into the differential equation for $J^\pdag_{t't}(B)$ in $O(J^4)$.
We are therefore left with identifying the terms with structure
$\vec S \cdot \vec s^\pdag_{t't}$ in the normal--ordered commutators
\bea
C_{3'}&=&[\sum_{u',u} \eta^{(1)}_{u'u}\, \vec S \cdot \vec s^\pdag_{u'u}\, , \\
&&\qquad i\sum_{v',v,w',w} K^\pdag_{v'v,w'w}\,
:\vec S\cdot (\vec s^\pdag_{v'v} \times \vec s^\pdag_{w'w}):] \nn 
\eea
and
\bea
C_{3''}&=&[i\sum_{v',v,w',w} \eta^{(2)}_{v'v,w'w}\,
:\vec S\cdot (\vec s^\pdag_{v'v} \times \vec s^\pdag_{w'w}): \, ,\nn \\
&&\qquad \sum_{u',u} J^\pdag_{u'u}\, \vec S \cdot \vec s^\pdag_{u'u}] \ . 
\eea
We can combine these two contributions into
\bea
C_3&=& \sum (\eps_{u'}-\eps_u-\eps_{v'}+\eps_v-\eps_{w'}+\eps_w)
\,J^\pdag_{u'u} K^\pdag_{v'v,w'w} \nn \\
&&\times\: i\,[\vec S \cdot \vec s^\pdag_{u'u}\, , \,
:\vec S\cdot (\vec s^\pdag_{v'v} \times \vec s^\pdag_{w'w}):] \ .
\eea
The commutator $[\vec S \cdot \vec s^\pdag_{u'u}\, , \,
:\vec S\cdot (\vec s^\pdag_{v'v} \times \vec s^\pdag_{w'w}):]$ is worked out
in the appendix and given in (\ref{app_comm2}). We only need the 
contribution with structure $\vec S \cdot \vec s^\pdag_{t't}$ and inserting
yields
\bea
C_3&=&\frac{1}{2}\:\sum (2\eps_u-2\eps_{u'}+\eps_t-\eps_{t'})
 \: \vec S \cdot \vec s^\pdag_{t't}
\label{C3} \\
&&\times J^\pdag_{u'u} (K^\pdag_{u'u,t't}-K^\pdag_{t't,u'u}) \nn \\
&&\times \big(n(u')\,(1-n(u))+n(u)\,(1-n(u'))\big)
 \nn \ .
\eea
We have now derived all the necessary terms to write down the flow equations
to the desired order.

\subsubsection{Flow Equations}
We include the term (\ref{C3}) in the differential equation (\ref{eq_J1loop})
and find the following set of flow equations that completely determine the
flow of the Hamiltonian $H(B)$ up to the resp.\ order in~$J$
\bea
\frac{dJ^\pdag_{t't}}{dB}&=&-(\epsilon^\pdag_{t'}-\epsilon^\pdag_t)^2
J^\pdag_{t't}
\label{eq_J2loop} \\
&&+\sum_v  (\epsilon^\pdag_{t'}+\epsilon^\pdag_t-2\epsilon^\pdag_v)
J^\pdag_{t'v}   J^\pdag_{vt} \,(n(v)-1/2) \nn \\
&&+\frac{1}{2}\: \sum_{u',u} (2\eps_u-2\eps_{u'}+\eps_t-\eps_{t'}) 
\nn \\
&&\quad\times J^\pdag_{u'u} (K^\pdag_{u'u,t't}-K^\pdag_{t't,u'u})
\nn \\
&&\quad\times \big(n(u')\,(1-n(u))+n(u)\,(1-n(u'))\big) \nn \\
&&+O(J^4) \nn \\
\frac{dK^\pdag_{t't,u'u}}{dB} &=&
-(\epsilon^\pdag_{t'}+\epsilon^\pdag_{u'}-\epsilon^\pdag_t-\epsilon^\pdag_u)^2\,
K^\pdag_{t't,u'u}
\label{eq_K2loop} \\
&&-(\epsilon^\pdag_{u'}-\epsilon^\pdag_u)
J^\pdag_{t't}   J^\pdag_{u'u} \nn \\
&& +O(J^3) \nn
\eea
In order to complete our analysis we finally need to show that the
terms $V_{t't}(B)$ from (\ref{potscat}) indeed vanish to the required order.
This is straightforward by solving (\ref{eq_J2loop}) in linear order
\beq
J^\pdag_{t't}(B)=(J^\pdag_{t't}(B=0)+O(J^2))\,e^{-B(\eps_{t'}-\eps_t)^2} 
\label{J_0loop}
\eeq
and inserting in (\ref{feq_V})
\bea
\frac{dV^\pdag_{t't}}{dB} &=&\frac{3}{16}\, 
\sum_v  J^\pdag_{t'v}(B=0)   J^\pdag_{vt}(B=0) \\
&&\times (\epsilon^\pdag_{t'}+\epsilon^\pdag_t-2\epsilon^\pdag_v)
\:e^{-B((\eps_{t'}-\eps_v)^2+(\eps_v-\eps_t)^2)} \nn \\
&& + O(J^3) \nn
\eea
Since we assume a linear dispersion relation and initial coupling constants 
$J_{t't}(B=0)$ that are constant (at least in the low--energy limit) we only
need to know 
\beq
\sum_v  (\epsilon^\pdag_{t'}+\epsilon^\pdag_t-2\epsilon^\pdag_v)
\:e^{-B((\eps_{t'}-\eps_v)^2+(\eps_v-\eps_t)^2)}=0
\eeq
which holds up to possible band edge effects that do no influence the
universal low--energy physics for $T_{\rm K}\ll D$. Therefore
\beq
\frac{dV^\pdag_{t't}}{dB}=O(J^3)
\label{eq_V}
\eeq
and the $V_{t't}$--terms do not influence the flow equations 
(\ref{eq_J2loop}) and (\ref{eq_K2loop}) in the orders that we are interested
in. In fact, a more detailed analysis shows that even the $O(J^3)$--terms
on the rhs of (\ref{eq_V}) vanish; the first nonvanishing contributions
arise in order~$J^4$.  

Eqs.~(\ref{eq_J2loop}) and (\ref{eq_K2loop}) constitute the main computational
results of this paper and will now be analyzed in various settings. Notice
that they apply for a general spin--1/2 multi--lead Kondo model where we did
not yet have to specify whether the system is equilibrium or not. The
derivation of these equations followed from a straightforward application of
the canonical flow equation framework as an expansion in powers of the flowing 
coupling constants. Also notice that the linear terms in~(\ref{eq_J2loop}) 
and (\ref{eq_K2loop}) generate the canonical exponential decay of the
coupling constants (\ref{J_0loop}) that make the Hamiltonian $H(B)$
increasingly diagonal in energy space (see Fig.~\ref{Fig_FEQvsRG2})
with the identification $\Lambda_{\rm feq}=B^{-1/2}$.   

\subsection{Equilibrium Model}
As a consistency check and in order to gain some insight into the system
of flow equations we now first analyze the equilibrium Kondo model
at zero temperature $T=0$.
Here the index~$t$ labelling the conduction band electrons in 
(\ref{H_t}) only consists of the electron momentum and we have one
exchange coupling~$J$. We use the following approximate parametrization
of the running couplings as a function of~$B$
\beq
\rho\,J_{p'p}(B) = g(B)\,e^{-B(\eps_{p'}-\eps_p)^2}
\label{g_IR}
\eeq
where $g(B)$ is determined from the flow in the 
IR--limit $\eps_{p'}=\eps_p=0$
(we choose $\eps_F=0$). This parametrization is asymptotically correct in
the IR--limit and we can identify $g(B)$ with the dimensionless running
coupling constant of the conventional scaling approach.

Inserting the parametrization (\ref{g_IR}) into the flow equation 
(\ref{eq_J2loop}) for $\eps_{p'}=\eps_p=0$ 
\bea
\frac{dg}{dB}&=&g^2\int_{-\infty}^\infty d\eps\:
e^{-2B\epsilon^2}\,(-2\eps)\,(n(\eps)-1/2)
\nn \\
&&+\int_{-\infty}^0 d\epsilon' \int_0^\infty d\eps\,2(\eps-\epsilon')\,
g\,e^{-B(\epsilon'-\eps)^2} \nn \\
&&\qquad\times (K^\pdag_{\epsilon' \epsilon,00}-K^\pdag_{00,\epsilon'
  \epsilon}) 
\label{eq_g_equ}
\eea
where we have insert the zero temperature occupation numbers
\beq
n(\eps)=\Theta(-\eps) \ .
\eeq
Eq.~(\ref{eq_g_equ}) is integrated starting from $B=D^{-2}$ where 
$D$ is the conduction band width (UV--cutoff). We also need to
the flow of the $K$--terms that are initially not present in the
Hamiltonian but generated during the flow: from (\ref{eq_K2loop})
we deduce
\bea
\frac{dK_{\epsilon' \epsilon,00}}{dB} &=& 0 \\
\frac{dK_{00,\epsilon' \epsilon}}{dB} &=&
-(\epsilon'-\epsilon)^2\,K_{00,\epsilon' \epsilon}
\label{eq_K_equ} \\
&&-g^2\,e^{-B(\epsilon'-\eps)^2}\,(\epsilon'-\eps) \nn
\eea
Eq.~(\ref{eq_K_equ}) can be solved easily 
\beq
K_{00,\epsilon' \epsilon}(B)=-(\epsilon'-\eps)\,e^{-B(\epsilon'-\eps)^2}\:
\int_0^B dB'\,g^2(B') \ .
\label{eq_K_equ1}
\eeq
From the scaling equation in second order for $g(B)$ one obtains the
well--known slow logarithmic increase during the flow, therefore the
integral in (\ref{eq_K_equ1}) is dominated by large values of~$B'$
and one can write
\beq
K_{00,\epsilon' \epsilon}(B)=-(\epsilon'-\eps)\,e^{-B(\epsilon'-\eps)^2}\:
B\,(g^2(B)+O(g^3)) \ .
\eeq
Putting everything together in (\ref{eq_g_equ}) and performing the
remaining integrations we obtain
\beq
\frac{dg}{dB}=\frac{g^2}{2B}-\frac{g^3}{4B} +\frac{O(g^4)}{B} \ .
\label{RG_gB}
\eeq
In terms of the scaling parameter $\Lambda_{\rm feq}=B^{-1/2}$ with
dimension energy one can alternatively write
\beq
\frac{dg}{d\ln\Lambda_{\rm feq}}=-\beta^{\rm (eq)}(g) 
\eeq
with the correct equilibrium $\beta$--function to two--loop order \cite{Hewson}
\beq
\beta^{\rm (eq)}(g)=g^2-\frac{1}{2}\,g^3+O(g^4) \ .
\label{betafunction}
\eeq
As expected
for the equilibrium model it therefore makes no difference whether one
derives the IR--behavior with respect to the conventional scaling parameter
$\Lambda_{\rm RG}$ or the flow equation scaling parameter $\Lambda_{\rm feq}$
(see Fig.~\ref{Fig_FEQvsRG2}).
In the next section we will explore how these results change in
the non--equilibrium situation.

\subsection{Non--Equilibrium Model (general parameters)}
For the non--equilibrium setting with voltage bias~$V$ between the left
and right lead the label~$t$ in (\ref{H_t}) is a multi--index $t=(a,p)$ 
labelling both left/right lead ($a=l,r$) and electron momentum~$p$.
We choose the Fermi energy of the left lead to be~$+V/2$ and of the
right lead to be~$-V/2$, therefore we have a current flowing from
left to right. For now we will only look at zero temperature~$T=0$.

In our flow equations (\ref{eq_J2loop}) and (\ref{eq_K2loop}) we
have to differentiate between the various scattering processes between 
the leads:
\bea
\lefteqn{
\frac{dJ^\pdag_{(a'p')(ap)}}{dB}=-(\epsilon^\pdag_{p'}-\epsilon^\pdag_p)^2
J^\pdag_{(a'p')(ap)} }
\label{neq_J2loop} \\
&&+\sum_{b=l,r} \sum_q  (\epsilon^\pdag_{p'}+\epsilon^\pdag_p-2\epsilon^\pdag_q)
J^\pdag_{(a'p')(bq)}   J^\pdag_{(bq)(ap)} \nn \\
&&\qquad\times (n_b(q)-1/2) \nn \\
&&+\frac{1}{2}\: \sum_{b',b=l,r} \sum_{q',q} (2\eps_q-2\eps_{q'}+\eps_p-\eps_{p'}) 
\,J^\pdag_{(b'q')(bq)}
\nn \\
&&\qquad\times  (K^\pdag_{(b'q')(bq),(a'p')(ap)}
-K^\pdag_{(a'p')(ap),(b'q')(bq)}) \nn \\
&&\qquad\times \big(n_{b'}(q')\,(1-n_b(q))+n_{b}(q)\,(1-n_{b'}(q'))\big)
\nn \\
&&+O(J^4) \nn 
\eea
for $a',a=l,r$ and
\bea
\lefteqn{
\frac{dK^\pdag_{(a'p')(ap),(b'q')(bq)}}{dB} }
\label{neq_K2loop} \\
 &=&
-(\epsilon^\pdag_{p'}+\epsilon^\pdag_{q'}-\epsilon^\pdag_p-\epsilon^\pdag_q)^2 \,
K^\pdag_{(a'p')(ap),(b'q')(bq)}
\nn \\
&&-(\epsilon^\pdag_{q'}-\epsilon^\pdag_q)
J^\pdag_{(a'p')(ap)}   J^\pdag_{(b'q')(bq)} \nn \\
&& +O(J^3) \nn
\eea
for $b',b,a',a=l,r$. The effect of the voltage bias enters here
only through the different ground state expectation values in
the left and right lead
\beq
n_l(\eps)=\Theta(-\eps+V/2) \quad , \quad n_r(\eps)=\Theta(-\eps-V/2) \ . 
\eeq

\subsection{Non--Equilibrium Model ($J_{lr}^2=J_{ll} J_{rr}$)}
Eqs.~(\ref{neq_J2loop}) and (\ref{neq_K2loop}) can be simplified
considerably if the Kondo model can be derived from an underlying
Anderson single impurity model. As explained in Sect.~\ref{ch_Kondomodel},
the coupling constants then fulfill the relations 
$J_{lr}^2=J_{ll} J_{rr}$ and $J_{ll}/J_{rr}=\Gamma_l/\Gamma_r=r$.
Notice that the impurity physics is invariant under the exchange $r\,\leftrightarrow 1/r$
since this just amounts to an exchange of the two leads.
One can verify easily that the impurity spin only couples
to the following linear combination of left and right lead
fermion operators
\beq
f_{p\alpha} \stackrel{\rm def}{=} 
\frac{1}{\sqrt{1+r}}\: c_{rp\alpha} + \frac{1}{\sqrt{1+r^{-1}}}\: c_{lp\alpha} \ ,
\eeq
which obey the usual anticommuation relations
$\{f^\pdag_{p\alpha},f^\dagger_{p'\beta}\}
=\delta^\pdag_{pp'}\delta^\pdag_{\alpha\beta}$. The flowing Hamiltonian
then takes the form
\bea
H(B)&=&\sum_{p,\alpha} \epsilon^\pdag_p\,f^\dag_{p\alpha} f^\pdag_{p\alpha}
+\sum_{p',p} J^\pdag_{p'p}(B)\, \vec S \cdot \vec s^\pdag_{p'p} 
\label{HBa} \\
&&+i\sum_{p',p,q',q} K^\pdag_{p'p,q'q}(B)\,
:\vec S\cdot (\vec s^\pdag_{p'p} \times \vec s^\pdag_{q'q}): \nn \ ,
\eea 
with the conduction band spin operators 
$\vec s^\pdag_{p'p}=\frac{1}{2}\sum_{\alpha,\beta} f^\dag_{p'\alpha}
\vec\sigma^\pdag_{\alpha\beta} f^\pdag_{p\beta}$ defined for the
$f$-operators. Eqs.~(\ref{neq_J2loop}) and (\ref{neq_K2loop})
simplify to the following form where the external leads do not
appear explicitly:
\bea
\lefteqn{
\frac{dJ^\pdag_{p'p}}{dB}=-(\epsilon^\pdag_{p'}-\epsilon^\pdag_p)^2
J^\pdag_{p'p} }
\label{neq_J2loopa} \\
&&+\sum_q  (\epsilon^\pdag_{p'}+\epsilon^\pdag_p-2\epsilon^\pdag_q)
J^\pdag_{p'q}   J^\pdag_{qp}\, (n^\pdag_f(q)-1/2) \nn \\
&&+\frac{1}{2}\: \sum_{q',q} (2\eps_q-2\eps_{q'}+\eps_p-\eps_{p'}) 
\,J^\pdag_{q'q}
\nn \\
&&\qquad\times  (K^\pdag_{q'q,p'p}
-K^\pdag_{p'p,q'q}) \nn \\
&&\qquad\times \big(n^\pdag_{f}(q')\,(1-n^\pdag_f(q))
+n^\pdag_{f}(q)\,(1-n^\pdag_{f}(q'))\big)
\nn \\
&&+O(J^4) \nn 
\eea
\bea
\frac{dK^\pdag_{p'p,q'q}}{dB} 
 &=&
-(\epsilon^\pdag_{p'}+\epsilon^\pdag_{q'}-\epsilon^\pdag_p-\epsilon^\pdag_q)^2 \,
K^\pdag_{p'p,q'q}
\label{neq_K2loopa} \\
&&-(\epsilon^\pdag_{q'}-\epsilon^\pdag_q)
J^\pdag_{p'p}   J^\pdag_{q'q} \nn \\
&& +O(J^3) \nn
\eea
Here we have the initial condition $J_{p'p}(B=0)=J_{ll}+J_{rr}$ and
the zero temperature Fermi distribution function for the $f$-operators
\bea
n_f(p)&=&
\langle f^\dagger_{p\alpha} f^\pdag_{p\alpha} \rangle \nn \\
&=&\frac{1}{1+r}\,n_r(p) + \frac{1}{1+r^{-1}}\,n_l(p) \\
&=&\left\{ \begin{array}{cl}
0 &\quad \eps_p>\frac{V}{2} \\
\frac{\displaystyle 1}{\displaystyle 1+r^{-1}} &\quad |\eps_p|\leqslant \frac{V}{2} \\
1 &\quad \eps_p<-\frac{V}{2} \ .
\end{array} \right. 
\label{occnonequ}
\eea
In the sequel we will restrict ourselves to the analysis of 
(\ref{neq_J2loopa}) and (\ref{neq_K2loopa}), that is we only investigate
a Kondo model that can be realized from an underlying single Anderson impurity model.

\section{Scaling Picture of the Kondo Model with Voltage Bias}
\subsection{IR--parametrization}
Eqs.~(\ref{neq_J2loopa}) and (\ref{neq_K2loopa}) contain the full
information about the Hamiltonian flow to the resp.~order in the
coupling constant. We will analyze these equations both
numerically and analytically to understand the scaling behavior
of the non--equilibrium Kondo model. In order to get some first
insights, we first perform an analytical analysis that is based
on some additional approximations. However, we will later see
that these approximations are justified with very good accuracy
by comparison with the exact numerical solution.

The flow equation differential equations are connected with
conventional scaling equations by using an approximation of
the following form
\beq
\rho J^\pdag_{p'p}(B)=u^\pdag_{\overline{p'p}}(B)\,e^{-B(\eps_{p'}-\eps_p)^2} \ ,
\eeq
which parametrizes the couplings in terms of the dimensionless running coupling
constants $u^\pdag_{\overline{p'p}}(B)$ on an averaged energy scale 
$\eps_{\overline{p'p}}\stackrel{\rm def}{=}(\eps_{p'}+\eps_p)/2$. (In fact
one can choose any energy in $[{\rm min}(\eps_p,\eps_{p'}), {\rm max}(\eps_p,\eps_{p'}]$
with excellent accuracy.) This ansatz solves the linear part of the flow equation
(\ref{neq_J2loopa}) and allows us to perform the summations over $q$
in closed form. One finds
\bea
\frac{du_p}{dB}&=&\frac{u_p^2}{2B}\,\Big(\frac{1}{1+r}\,e^{-2B(-\eps_p-V/2)^2} \\
&&\qquad + \frac{1}{1+r^{-1}}\,e^{-2B(-\eps_p+V/2)^2} \Big) \nn \\
&&-2\int d\eps_{q'}\,d\eps_q\: (\eps_{q'}-\eps_q)^2\,u_q\,e^{-2B(\eps_{q'}-\eps_q)^2} \nn \\
&&\qquad \times
\int_0^B dB'\,u_p(B')\,u_q(B') \nn \\
&&\qquad \times n^\pdag_{f}(q')\,(1-n^\pdag_f(q)) \ . \nn
\eea
In the cubic term a whole range of values $q',q$ contributes to the integral
where the product of the Fermi functions is nonzero. Again with very good
accuracy we can replace the couplings $u_{q'}, u_q$ under the integral
by an average over the window $[-V/2,V/2]$ that is responsible for
transport,
\beq
u_t(B) \stackrel{\rm def}{=} \frac{1}{V}\,\int_{-V/2}^{V/2} d\epsilon_q \, u_q(B) \ .
\label{def_ut}
\eeq
The integrals over $d\eps_{q'}$ and $d\eps_q$ can then be performed in closed
form and one arrives at the following differential equation
\bea
\lefteqn{
\frac{du_p}{dB}=} 
\label{eq_up} \\
&&=\frac{u_p^2(B)}{2B}\Big(\frac{1}{1+r}\,e^{-2B(-\eps_p-V/2)^2} \nn \\
&&\qquad\qquad\qquad + \frac{1}{1+r^{-1}}\,e^{-2B(-\eps_p+V/2)^2} \Big) \nn \\
&&-\frac{u_t(B)}{4B^2}\,\int_0^B dB'\,u_p(B')\,u_t(B') \nn \\
&&\times\Big(\frac{r+r^{-1}}{(1+r)(1+r^{-1})} \nn \\
&&+\frac{1}{(1+r)(1+r^{-1})}\big(2e^{-2BV^2}+\sqrt{2\pi B}V\,{\rm erf}(\sqrt{2B}V)\big)
\Big) \nn
\eea   
We first use this to derive a closed equation for 
$u_t(B)$ from (\ref{def_ut}) by averaging over~$p$
\bea
\lefteqn{
\frac{du_t}{dB}=
\frac{u_t^2(B)}{2B}\,\frac{\sqrt{\pi}}{\sqrt{8B}V}\,
{\rm erf}(\sqrt{2B}V) } 
\label{eq_ut} \\
&&-\frac{u_t(B)}{4B^2}\,\int_0^B dB'\,u^2_t(B') \nn \\
&&\times\Big(\frac{r+r^{-1}}{(1+r)(1+r^{-1})} \nn \\
&&+\frac{1}{(1+r)(1+r^{-1})}\big(2e^{-2BV^2}+\sqrt{2\pi B}V\,{\rm erf}(\sqrt{2B}V)\big)
\Big) \nn
\eea  
Of particular importance are also the running coupling constants at the left and
right Fermi surfaces
\beq
u_l(B)\stackrel{\rm def}{=} u_{\eps_p=V/2}(B) \ ,\quad 
u_r(B)\stackrel{\rm def}{=} u_{\eps_p=-V/2}(B) \ ,
\eeq
which will turn out to be relevant for the behavior of the quasiparticle
resonances (Kondo peaks) and which determine the phase diagram since they 
correspond to extremal values of $u_p(B=\infty)$ as a function of~$B$ 
(see Sect.~\ref{ch_numerics}). Their flow equations follow immediately
from (\ref{eq_up}):
\bea
\lefteqn{
\frac{du_l}{dB}=  
\frac{u_l^2(B)}{2B}\Big(\frac{1}{1+r^{-1}}
+ \frac{1}{1+r}\,e^{-2BV^2} \Big) } 
\label{eq_ul} \\
&&-\frac{u_t(B)}{4B^2}\,\int_0^B dB'\,u_l(B')\,u_t(B') \nn \\
&&\times\Big(\frac{r+r^{-1}}{(1+r)(1+r^{-1})} \nn \\
&&+\frac{1}{(1+r)(1+r^{-1})}\big(2e^{-2BV^2}+\sqrt{2\pi B}V\,{\rm erf}(\sqrt{2B}V)\big)
\Big) \nn
\eea 
\bea
\lefteqn{
\frac{du_r}{dB}=  
\frac{u_r^2(B)}{2B}\Big(\frac{1}{1+r}
+ \frac{1}{1+r^{-1}}\,e^{-2BV^2} \Big) } 
\label{eq_ur} \\
&&-\frac{u_t(B)}{4B^2}\,\int_0^B dB'\,u_r(B')\,u_t(B') \nn \\
&&\times\Big(\frac{r+r^{-1}}{(1+r)(1+r^{-1})} \nn \\
&&+\frac{1}{(1+r)(1+r^{-1})}\big(2e^{-2BV^2}+\sqrt{2\pi B}V\,{\rm erf}(\sqrt{2B}V)\big)
\Big) \nn
\eea 
The three differential equations (\ref{eq_ut}), (\ref{eq_ul}) and (\ref{eq_ur})
are one of the key results of this work and allow us to describe and understand the scaling
behavior of the non--equilibrium Kondo model.\cite{footnote_us}

\subsection{Scaling Analysis}
We will first analyze (\ref{eq_ut}), (\ref{eq_ul}) and (\ref{eq_ur}) in the
initial phase of the flow where $\Lambda_{\rm feq}\gg V$. As expected, all
three differential equations coincide and one finds for $a=l,r,t$:
\beq
\frac{du_a}{d\Lambda_{\rm feq}}=-\frac{u_a^2}{\Lambda_{\rm feq}}
+\frac{u_a^3}{2\Lambda_{\rm feq}} +\frac{O(u_a^4)}{\Lambda_{\rm feq}} \ .
\label{eq_equ_u}
\eeq
Here we have used the approximation $\int_0^B dB'\,u^2_a(B')=B\,u^2_a(B)$,
which holds plus correction terms in higher order due to the slow logarithmic
flow of $u_a(B)$. The initial behavior of the running coupling constants
is therefore determined by the equilibrium $\beta$-function (\ref{betafunction})
as expected. With our initial condition $u(B=0)=g_l+g_r$ this also agrees
exactly with the scaling equations (\ref{eq_Glazman99}) from the analysis
of Kaminski {\it et al.\/}\cite{Glazman99} (with the identification 
$\Lambda_{\rm RG}=\Lambda_{\rm feq}$).

Once the flow parameter is smaller than the voltage bias, 
$\Lambda_{\rm feq}\ll V$, the scaling equations take a different
structure. The effective transport coupling obeys 
\beq
\frac{du_t}{d\Lambda_{\rm feq}}=u_t^3\,\frac{V}{\Lambda_{\rm feq}^2}\,
\sqrt{\frac{\pi}{2}}\,\frac{1}{(1+r)(1+r^{-1})} \ ,
\label{eq_uta}
\eeq
and the couplings at the left and right Fermi surfaces, resp.,
\bea
\frac{du_l}{d\Lambda_{\rm feq}}&=&-\frac{1}{1+r^{-1}}\,\frac{u_l^2}{\Lambda_{\rm feq}}
\label{eq_ula} \\
&&+u_l\,u_t^2\,\frac{V}{\Lambda_{\rm feq}^2}\,
\sqrt{\frac{\pi}{2}}\,\frac{1}{(1+r)(1+r^{-1})} \nn \\
\frac{du_r}{d\Lambda_{\rm feq}}&=&-\frac{1}{1+r}\,\frac{u_r^2}{\Lambda_{\rm feq}} 
\label{eq_ura} \\
&&+u_r\,u_t^2\,\frac{V}{\Lambda_{\rm feq}^2}\,
\sqrt{\frac{\pi}{2}}\,\frac{1}{(1+r)(1+r^{-1})} \ . \nn 
\eea
One notices that the strong-coupling growth of the average transport
coupling~$u_t$ stops when $\Lambda_{\rm feq}\sim V$ due to the energy
difference of the left and right Fermi surfaces. On the other hand,
the couplings at the left and right Fermi surfaces, $u_l$ and $u_r$, still 
exhibit the typical Kondoesque strong-coupling behavior in quadratic order 
on the rhs of (\ref{eq_ula}) and (\ref{eq_ura}). This reflects the 
inter-lead scattering processes from the left lead back into the left lead,
or likewise for the right lead: such processes still see a sharp Fermi
surface for all $\Lambda_{\rm feq}<V$. It is this obervation that has
led to the prediction of 2-channel Kondo physics in the Kondo model
with voltage bias based on a 1-loop calculation \cite{Coleman01}.

\begin{figure}[t]
\includegraphics[clip=true,width=8.0cm]{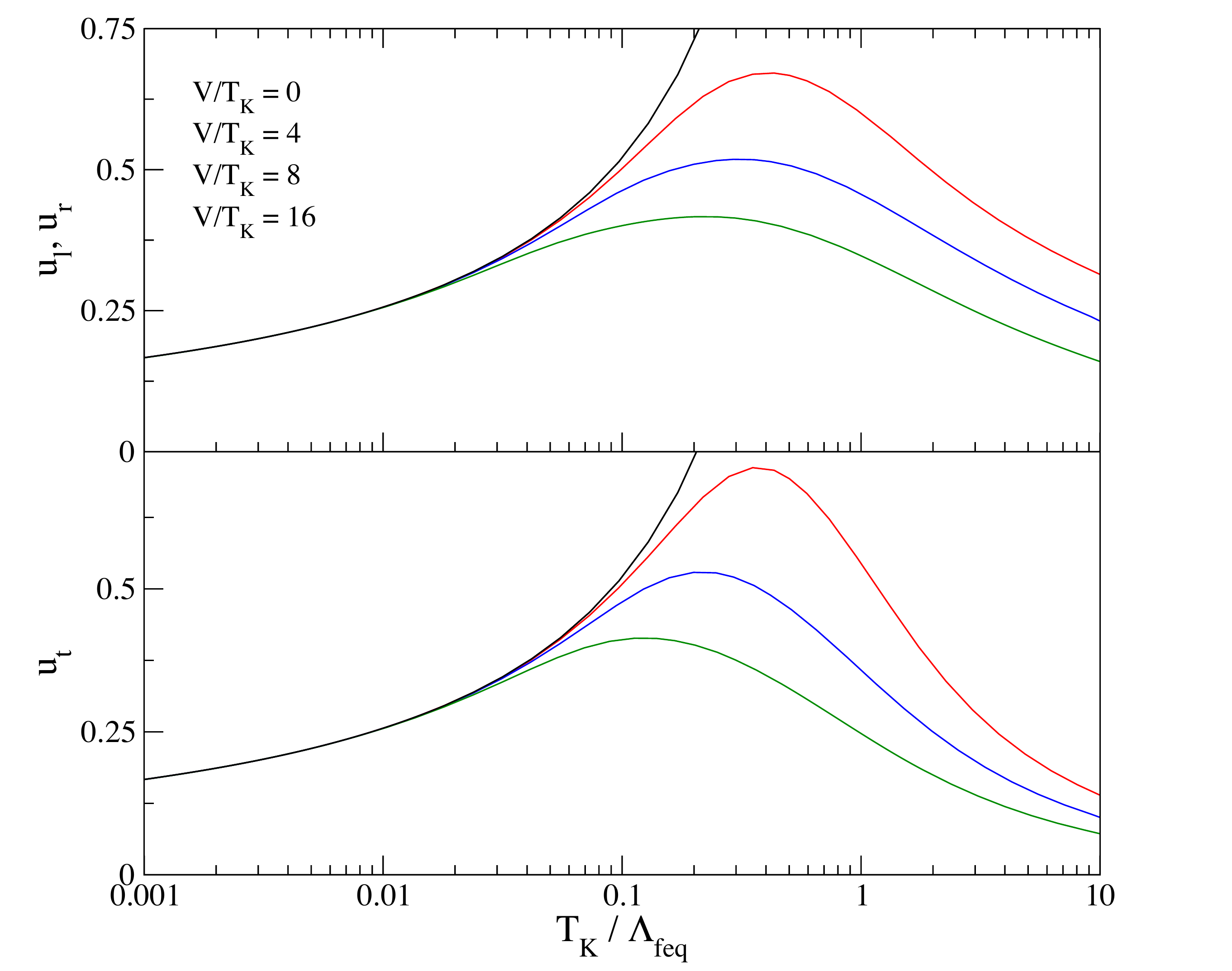}
\caption{Universal curves for the flow 
of $u_l$ (equal to $u_r$) and $u_t$ for symmetric Kondo dots ($r=1$). Results are shown
for various ratios $V/T_K$ labelling the curves from top to bottom.}
\label{fig_flows_of_u_sym}    
\end{figure}

\begin{figure}[t]
\includegraphics[clip=true,width=8.0cm]{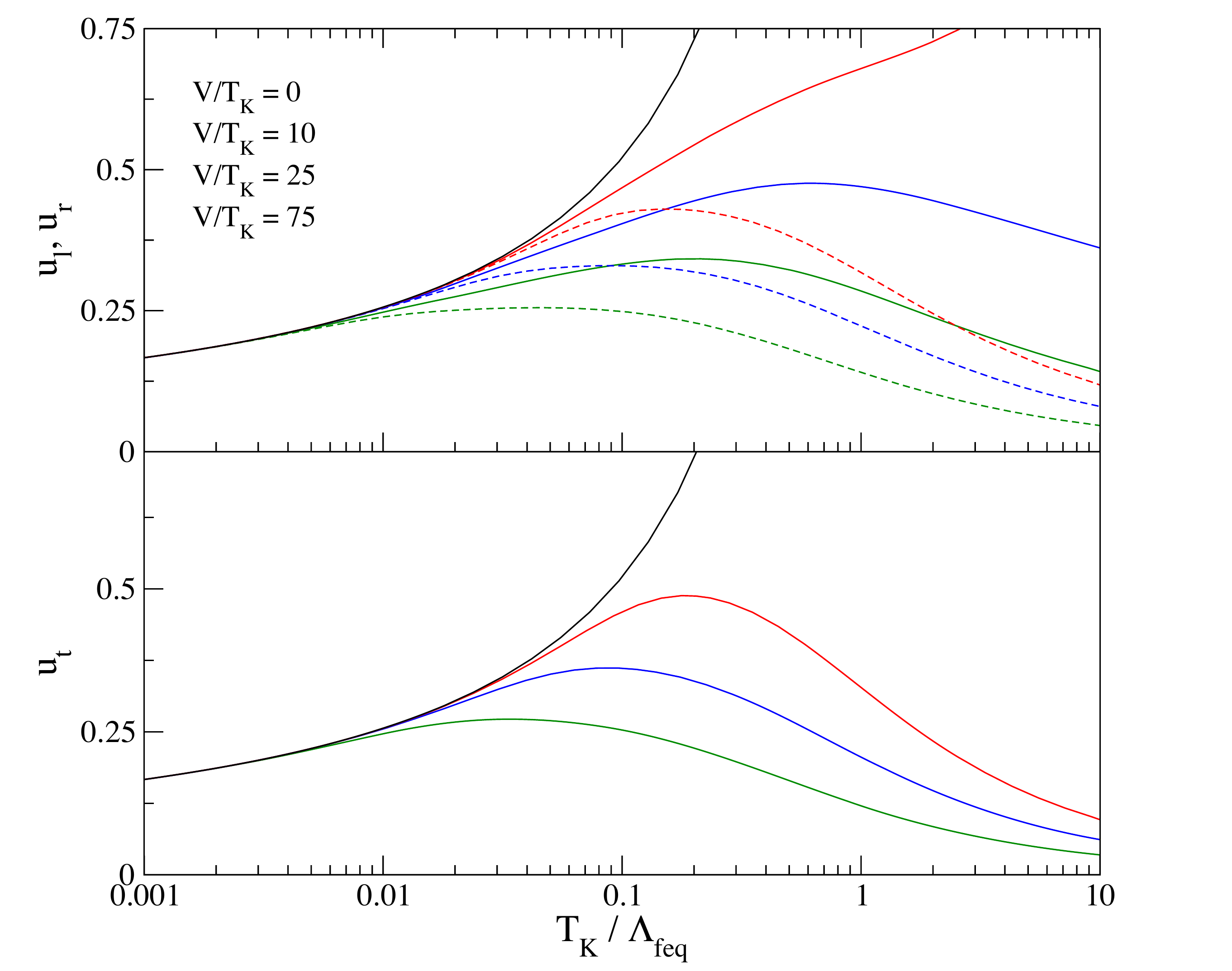}
\caption{Universal curves for the flow 
of $u_l$ (full lines), $u_r$ (dashed lines) and $u_t$ for asymmetrically 
coupled Kondo dots with $r=\Gamma_l/\Gamma_r=2$. Results are shown
for various ratios $V/T_K$ labelling the curves from top to bottom.
Notice that $u_l$ and $u_r$ coincide for $V=0$, therefore only $u_l$ is shown.}
\label{fig_flows_of_u_asym}    
\end{figure}

However, different from the experience in equilibrium models the 
third order terms in (\ref{eq_ula}) and (\ref{eq_ura}) can become
more important than the second order terms even for small coupling
constants: this is due to the more strongly growing $V/\Lambda^2_{\rm feq}$
terms as compared to the conventional $1/\Lambda_{\rm feq}$ behavior
in quadratic order in the infrared limit $\Lambda_{\rm feq}\rightarrow 0$.
Since the third order term has a positive sign, it counteracts
the second order term and can avoid the strong-coupling divergence.
Figs.~\ref{fig_flows_of_u_sym} and~\ref{fig_flows_of_u_asym} show 
the numerical solution of
(\ref{eq_ut}), (\ref{eq_ul}) and (\ref{eq_ur}) for symmetric ($r=1$) and
asymmetric Kondo dots ($r=2$). One can see that indeed all
couplings remain finite for sufficiently large voltage bias which
shows that there is no 2-channel strong-coupling divergence. 
 
Before proceeding with analyzing the resulting phase diagram in
more detail, we will first get some more analytical insights
into the solution of the scaling equations. The solution of
(\ref{eq_uta}) is straightforward
\beq
u_t(\Lambda_{\rm feq})=\frac{u_*}{\sqrt{1+\Gamma_{\rm rel}/\Lambda_{\rm feq}}} \ .
\eeq
Here
\beq
\Gamma_{\rm rel}\stackrel{\rm def}{=} \sqrt{2\pi}\,u_*^2\,V\,
\frac{1}{(1+r)(1+r^{-1})} \ ,
\label{def_Gammarel}
\eeq
where $u_*$ is the running coupling constant on the scale $\Lambda_{\rm feq}=V$
(since we can safely neglect $\Gamma_{\rm rel}/V$ in the weak-coupling regime).
From the solution of (\ref{eq_equ_u}) we then find
\beq
u_*=u(\Lambda_{\rm feq}=V)=\frac{1}{\ln(V/T_{\rm K})}
\eeq
with the Kondo temperature defined from the equilibrium model
\beq
T_{\rm K}=D\sqrt{g_l+g_r}\,e^{-1/(g_l+g_r)} \ .
\eeq
In order to understand the implication of the flow of~$u_t$ on
$u_l$ and $u_r$ it is convenient to first rewrite (\ref{eq_ula})
and (\ref{eq_ura}) in the following equivalent form:
\bea
\frac{du_l}{d\ln\Lambda_{\rm feq}}&=&-\frac{u_l^2}{1+r^{-1}}
+u_l\,\frac{d\ln u_t}{d\ln\Lambda_{\rm feq}} \\
\frac{du_r}{d\ln\Lambda_{\rm feq}}&=&-\frac{u_r^2}{1+r}
+u_r\,\frac{d\ln u_t}{d\ln\Lambda_{\rm feq}}  \ .
\eea
Since
\beq
\frac{d\ln u_t}{d\ln\Lambda_{\rm feq}}=\frac{1}{2}\,
\frac{\Gamma_{\rm rel}}{\Lambda_{\rm feq}+\Gamma_{\rm rel}} 
\label{eq_utb}
\eeq
this implies
\bea
\frac{du_l}{d\ln\Lambda_{\rm feq}}&=&u_l\left(-\frac{u_l}{1+r^{-1}}
+\frac{1}{2}\,\frac{\Gamma_{\rm rel}}{\Lambda_{\rm feq}+\Gamma_{\rm rel}}\right) 
\label{eq_ulb} \\
\frac{du_r}{d\ln\Lambda_{\rm feq}}&=&u_r\left(-\frac{u_r}{1+r}
+\frac{1}{2}\,\frac{\Gamma_{\rm rel}}{\Lambda_{\rm feq}+\Gamma_{\rm rel}}\right) \ .
\label{eq_urb}
\eea
These equations show a remarkable {\em transmutation} of the third order
terms in the running coupling constant into linear terms once 
$\Lambda_{\rm feq}\lesssim \Gamma_{\rm rel}$. This transmutation is
possible due to the $\Lambda_{\rm feq}^{-2}$-terms in (\ref{eq_ula})
and (\ref{eq_ura}) (or, equivalently, due to the $B^{-1/2}$-terms
in (\ref{eq_ul}) and (\ref{eq_ur}) as opposed to the equilibrium $B^{-1}$-terms).
For $\Lambda_{\rm feq}\gg\Gamma_{\rm rel}$ these linear order terms are
negligible and $u_l, u_r$ exhibit typical Kondoesque strong-coupling
scaling. In the later phase of the flow the couplings become
small again with 
$u_l(\Lambda_{\rm feq}), u_r(\Lambda_{\rm feq})\propto \Lambda^{1/2}_{\rm feq}$
if they are not already too large on the scale~$\Gamma_{\rm rel}$.

Let us analyze the meaning of this scale $\Gamma_{\rm rel}$.
We have already quoted the result from Kaminski {\it et al.\/}\cite{Glazman99}
that for $V\gg T_{\rm K}$ the current~$I$ is to leading order given by (\ref{eq_GV}) 
\beq
I=\frac{e^2}{h}\,\frac{3\pi^2}{2}\,u_*^2\,V\,\frac{1}{(1+r)(1+r^{-1})} \ .
\eeq
From (\ref{def_Gammarel}) we can see that $\Gamma_{\rm rel}\propto I$
with a propotionality factor independent of~$V$, $T_{\rm K}$ and~$r$.
It is therefore natural to associate $\Gamma_{\rm rel}$ with the 
spin decoherence rate generated by the shot noise proportional
to the current. The third order terms in the 
scaling equations therefore describe spin decoherence due to the
non--equilibrium current that cuts off the strong-coupling behavior
of the inter-lead scattering processes. This observation in our
Hamiltonian scaling framework is in agreement with the work
by Rosch {\it et al.\/}\cite{Rosch04b} that such non--equilibrium 
decoherence processes eliminate the 2-channel strong-coupling
divergence in the non--equilibrium Kondo model with $J_{lr}^2=J_{ll} J_{rr}$.

We will next use the flow equation analysis 
to study quantitatively the interplay between coherent (equilibrium)
strong-coupling processes and decoherence generated by the
non--equilibrium current in a systematic expansion in terms of
renormalized parameters. Before
proceeding along these lines, it is worthwile to point out that 
decoherence acts differently from temperature in the Kondo scaling
equations: Decoherence and strong-coupling physics are in competition
in (\ref{eq_ulb}) and (\ref{eq_urb}), whereas nonzero temperature~$T$ 
already eliminates the strong-coupling term itself, e.g.
\beq
\frac{dg}{d\ln\Lambda}=-g^2\,e^{-T/\Lambda} 
\label{eq_equKM_T}
\eeq
for the equilibrium Kondo model (\ref{g_IR}) in both the 
conventional scaling and the flow equation approach. We would also
like to mention the important question of how one can understand 
within the flow equation framework that the third order terms
in the scaling equations are really associated with {\em spin decoherence.}
We will postpone the answer to this question to the discussion of
the dynamic spin correlation function later in Sect.~\ref{ch_dynamicspin},
where we will work out this correspondence in detail.

\subsection{Weak-- and Strong--Coupling Regime}
From the scaling equation for $u_l$, $u_r$ and $u_t$ we can
now determine the scaling behavior of the non--equilibrium Kondo
model: that is we determine the regions in the parameter
space where all coupling constants remain small (weak-coupling
regime) or where at least one the coupling constants becomes
large (strong-coupling regime). Since only $u_l$ and $u_r$
can continue to grow below $\Lambda_{\rm feq}\approx V$
according to (\ref{eq_ulb}) and (\ref{eq_urb}) (as opposed to~$u_t$),
these regions are determined by the behavior of $u_l$ for 
$r=\Gamma_l/\Gamma_r>1$, or by the behavior of $u_r$ for $r<1$.
Because of the trivial exchange symmetry $r\leftrightarrow r^{-1}$,
we can focus on $r\geqslant 1$ below without loss of generality.

From (\ref{eq_ulb}) we can deduce an (approximate) condition
for $u_l(\Lambda_{\rm feq})$ remaining small during the entire flow
\beq
\frac{u_l(\Gamma_{\rm rel})}{1+r^{-1}} \lesssim \frac{1}{2} \ .
\eeq
In the notation of the original Kondo Hamiltonian (\ref{Kondo_nonequ})
this is equivalent to the inter-lead scattering processes being not
too large on the scale~$\Gamma_{\rm rel}$
\beq
\rho J_{ll}\lesssim \frac{1}{2} \ .
\eeq  
We can approximately rewrite this condition in terms of renormalized
quantities by explicitly integrating up the quadratic part of~(\ref{eq_ula}).
One finds
\beq
\frac{u_l(\Gamma_{\rm rel})}{1+r^{-1}}
=\frac{1}{\displaystyle \ln\left(\frac{V}{T_{\rm K}}\right)^{1+r^{-1}}
+\ln\left(\frac{\Gamma_{\rm rel}}{V}\right) }
\eeq
leading to
\beq
\left(\frac{V}{T_{\rm K}}\right)^{1+r^{-1}}
\:\frac{\sqrt{2\pi}}{
\ln^2 (V/T_{\rm K})}\:\frac{1}{(1+r)(1+r^{-1})} \gtrsim e^2 \ .
\label{eq_crossover}
\eeq
Comparison with the numerical solution
below shows that (\ref{eq_crossover}) can be used as an approxiate condition for 
the weak-coupling regime for $r\gtrsim 2$. Before proceeding with the
numerical solution of the full differential equation for $u_l$,
one can deduce some important analytical insights from 
(\ref{eq_crossover}): the ``critical" value of the voltage bias
$V/T_{\rm K}$ increases for increasing values of the asymmetry~$r$. 
For larger values of~$r$ this is due to the fact that decoherence
is proportional to the current, which is suppressed for asymmetric
coupling to the leads: $I/T_{\rm K}$ is maximum for $r=1$. The
third term on the lhs of (\ref{eq_crossover}) thus reflects the
$r$-dependence of the unitarity limit of the conductance~$G_u$
in (\ref{eq_Gu}).

\begin{figure}[t]
\includegraphics[clip=true,width=8.0cm]{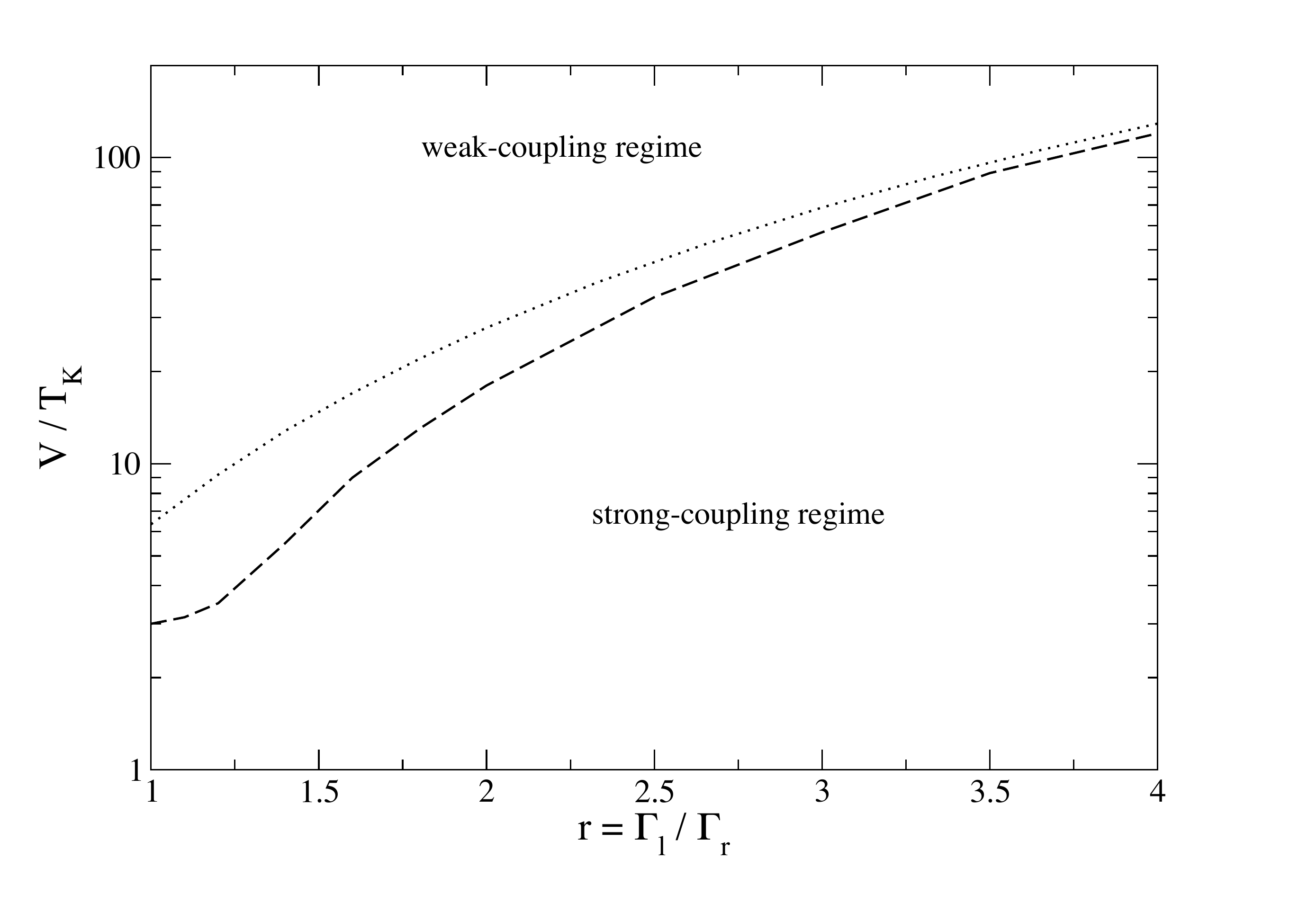}
\caption{Scaling picture of the non--equilibrium Kondo model as a function
of asymmetry $r=\Gamma_l/\Gamma_r$ and voltage bias. The dashed line
separates the weak-coupling regime from the strong-coupling regime
(deduced from the full numerical solution,
see text). The dotted line is the approximate analytical result
(\ref{eq_crossover}).}
\label{fig_phasediagram}    
\end{figure}

The above (approximate) analytical results are confirmed by the
exact numerical solution of the full differential equations 
(\ref{eq_ul}), (\ref{eq_ur}) and (\ref{eq_ut}) for the running
couplings $u_l, u_r$ and $u_t$. The definition of ``strong-coupling
regime" versus ``weak-coupling regime" is necessarily not unique
since we expect a smooth crossover between these regimes. The
definition used in our analysis is that in the weak-coupling regime
the couplings $u_l$ and $u_r$ remain
smaller than~0.75 during the entire flow, whereas in the 
strong-coupling regime at least one these coupling becomes larger than~0.75. 
Choosing a somehow different value than~0.75 (e.g.\ using~0.5 or 1.0 instead)
does actually hardly change the crossover line. The actual value of~0.75
is motivated by the observation that then the $T$-matrix reaches
the unitarity limit at the resp.\ Fermi surface 
in renormalized second order perturbation theory. This
indicates the breakdown of our perturbative expansion in the 
running coupling constant and therefore limits the region where our method
is reliable. The numerical results for this crossover line between
strong-coupling and weak-coupling regime are depicted in 
Fig.~\ref{fig_phasediagram}.

Fig.~\ref{fig_phasediagram} shows that the strong-coupling regime
extends to remarkably large values of the voltage bias for asymmetric
Kondo dots (as we have already qualitatively deduced from
(\ref{eq_crossover})). By comparison with the flow diagrams
in Figs.~\ref{fig_flows_of_u_sym} and~\ref{fig_flows_of_u_asym}
we can see that while there is no
2-channel Kondo physics (in the sense of a 2-channel strong-coupling
divergence), the crossover to the strong-coupling regime is
essentially given by 1-channel Kondo physics for asymmetric Kondo dots:
the couplings at the more strongly coupled Fermi surface dominate
the Kondo impurity physics.  

It should be emphasized again that one expects
a smooth crossover between weak-coupling and strong-coupling
regimes. While
this issue cannot be definitely answered using the weak--coupling tools here, there
is no physical reason to expect anything else than a smooth
crossover. Notice that for the equilibrium Kondo model at finite temperature
we do know that there is a smooth crossover between
weak-- and strong--coupling regimes.
In this context the question arises what physical properties actually
distinguish the strong-coupling regime from the weak-coupling regime.
We will address this question in detail later when we discuss the behavior
of observables: suitable observables behave quantitatively different in
these regimes and make the weak-coupling vs. strong-coupling 
distinction both useful and relevant
for the interpretation of experiments. For example 
in the strong-coupling regime the local
density of states remains nearly pinned to its Friedel value at the
more strongly coupled Fermi surface,
and it only decays once the voltage bias is well into the weak-coupling
regime.

\subsection{Expansion Schemes and Higher Orders}
We have already mentioned the remarkable {\em transmutation} of
a third order term in the running coupling constant into a linear
term, compare e.g. (\ref{eq_ula}) and (\ref{eq_ulb}) 
for~$u_l(\Lambda_{\rm feq})$. Decoherence therefore essentially
makes the running coupling irrelevant with a canonical scaling
dimension~$+1/2$ for $\Lambda_{\rm feq}\lesssim \Gamma_{\rm rel}$.  
This teaches an important lesson regarding the well-established
notion that higher order terms in the running coupling constant
cannot change the results of lower order calculations qualitatively
if the running coupling constant remains sufficiently small during
the flow (i.e., in the weak-coupling regime). It was exactly this
observation which eliminated the possibility of 2-channel Kondo
physics due to diverging inter-lead scattering processes. 

The technical reason for the transmutation in our non--equilibrium
model is the existence of the dimensionful parameter voltage bias~$V$,
which allows terms like $V/\Lambda_{\rm feq}^2$ alongside the
conventional $1/\Lambda_{\rm feq}$-terms in the scaling equations
(\ref{eq_uta}), (\ref{eq_ula}) and (\ref{eq_ura}). Therefore the
possibility of such a transmutation always arises whenever we have
dimensionful parameter in a scaling problem, which should be of
importance also for analyzing other problems and interpreting
the results of lower order calculations.
The unavoidable general conclusion is that the expansion order by order in the
running coupling constant is not a systematic expansion anymore,
unless we can show that no such ``transmutation'' occurs and changes the
results. Notice that for example
the equilibrium Kondo model at finite temperature contains the
dimensionful parameter temperature~$T$, which in fact also leads to
transmutation in its third order terms.\cite{footnote_finiteT} 
However, already the second order term
loses its strong-coupling behavior at sufficiently large temperature
according to (\ref{eq_equKM_T}), and therefore no qualitative changes
occur due to the third order terms.

It is worthwile to mention again the conceptual difference
between the flow equation framework and the conventional scaling analysis around
the two Fermi surfaces, compare Fig.~\ref{Fig_FEQvsRG}. For
$\Lambda_{\rm RG}\ll V$ the scaled Hamiltonian has no ``knowledge''
of the energy scale voltage bias anymore and therefore cannot
generate the $V/\Lambda^2_{\rm RG}$-terms in third order (2-loop order). 
On the other hand, the flow equation Hamiltonian still contains
all sufficiently energy diagonal scattering processes for
$\Lambda_{\rm feq}\ll V$ (compare Fig.~\ref{Fig_FEQvsRG2}), and
therefore ``knows'' about the energy voltage bias because this
is the energy window where scattering processes contributing
to the current are possible between the left and the right lead.  
It is exactly this observation which leads to the $V/\Lambda_{\rm feq}^2$-terms
in the flow equation scaling equations. 

In this context one should also address the effect
of higher order terms in the flow equation scaling equations. One
can verify that the most IR-singular behavior in fourth order
is $u^4\,V/\Lambda^2_{\rm feq}$, which is therefore smaller
than the third order terms in (\ref{eq_uta}-\ref{eq_ura}) in
the weak-coupling regime. In fifth order it seems possible to
have contributions that modify the $\sqrt{2\pi}$-proportionality
factor in our result for the relaxation rate (\ref{def_Gammarel}),
however, they cannot change the structure of (\ref{eq_ulb})
or (\ref{eq_urb}). More quantitative results would be a formidable
task equivalent to a 4-loop calculation. These observations 
indicate that the third order calculation presented here should
give reliable quantitative results in the entire non--equilibrium weak-coupling
regime.

\subsection{Numerical Solution}
\label{ch_numerics}
In the previous sections we have obtained analytical insights
into the competition of decoherence and strong-coupling behavior in the
non--equilibrium Kondo model. In order to do this we have used various
approximations starting from the full set of flow equations
(\ref{neq_J2loopa}) and (\ref{neq_K2loopa}). More accurate results,
which also allow us to check the accuracy of the approximations 
leading to (\ref{eq_uta}-\ref{eq_ura}), can be obtained by solving 
the full systems of differential equations. These solutions will
also be important for obtaining quantitative results for the behavior
of dynamical quantities later in Sect.~\ref{ch_observables}. The
numerical effort for solving (\ref{neq_J2loopa}) and (\ref{neq_K2loopa})
scales with $N^2$, where $N$ is the number of band states taken into
account for the numerical solution: details of the implementation 
of the numerics are contained in Appendix~\ref{app_numerics}. 

\begin{figure}[t]
\includegraphics[clip=true,width=8.0cm]{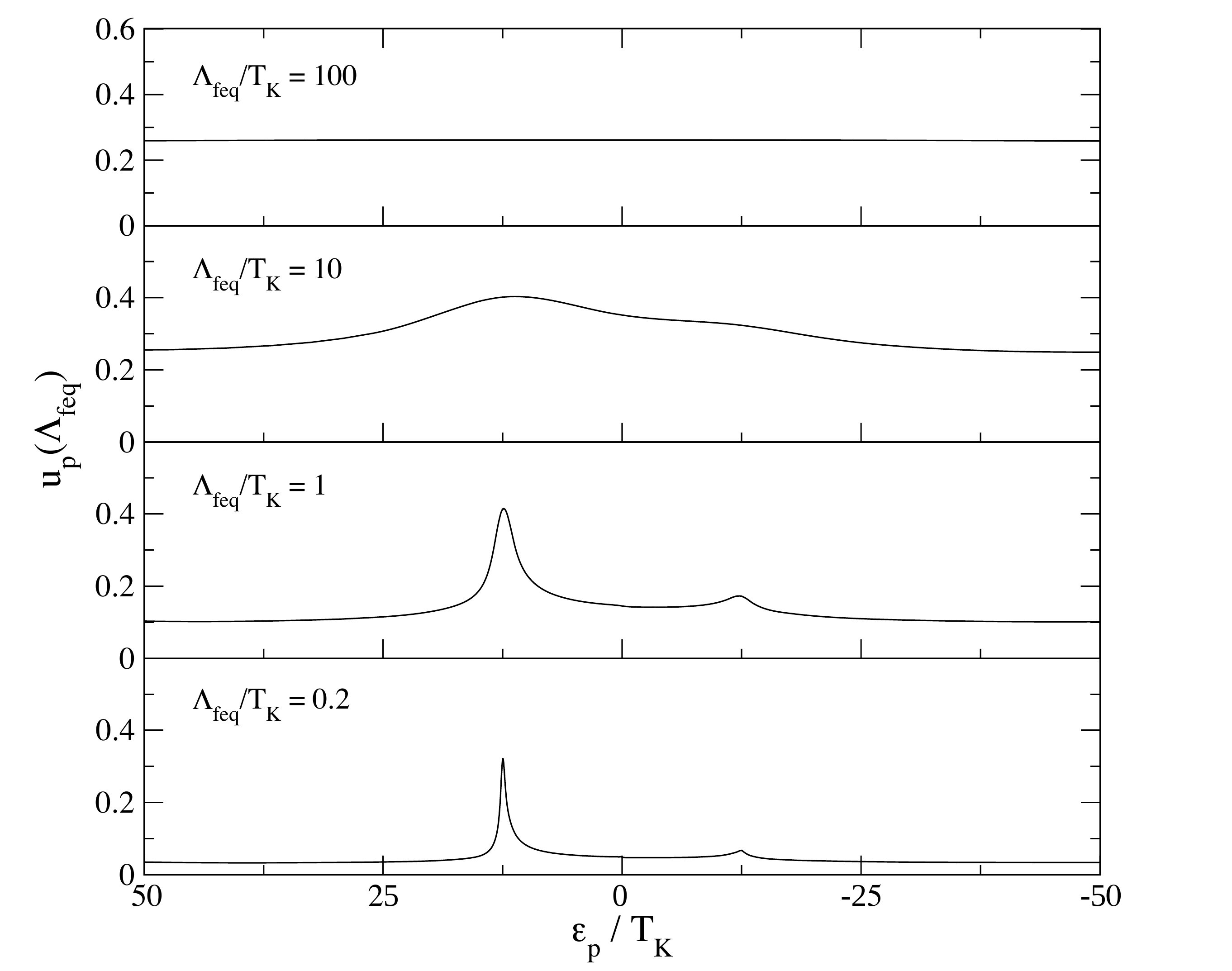}
\caption{Running coupling constants $u_p(\Lambda_{\rm feq})$ at various
points during the flow for an asymmetric Kondo dot with 
$r=\Gamma_l/\Gamma_r=2$ and $V/T_{\rm K}=25.0$. One observes
a much stronger buildup of strong-coupling behavior at the more
strongly coupled lead (Fermi surface at 
$\epsilon_p=12.5 T_{\rm K}$).}
\label{fig_flowup_asym}    
\end{figure}

Figs.~\ref{fig_flowup_sym} and \ref{fig_flowup_asym} depict the
flow of the coupling constants 
$u_p(\Lambda_{\rm feq})=\rho J_{pp}(\Lambda_{\rm feq})$ for a symmetric
and an asymmetric Kondo dot in the weak-coupling regime. One observes
the buildup of strong-coupling behavior at the left and right
Fermi surface, followed by decoherence effects that lead to a
decay of the coupling constants. In Fig.~\ref{fig_flowup_asym}
one can also notice the asymmetry in the buildup of the resonances
at the two Fermi surfaces due to the stronger effect of 
decoherence at the more weakly coupled lead. 
\begin{figure}[t]
\includegraphics[clip=true,width=8.0cm]{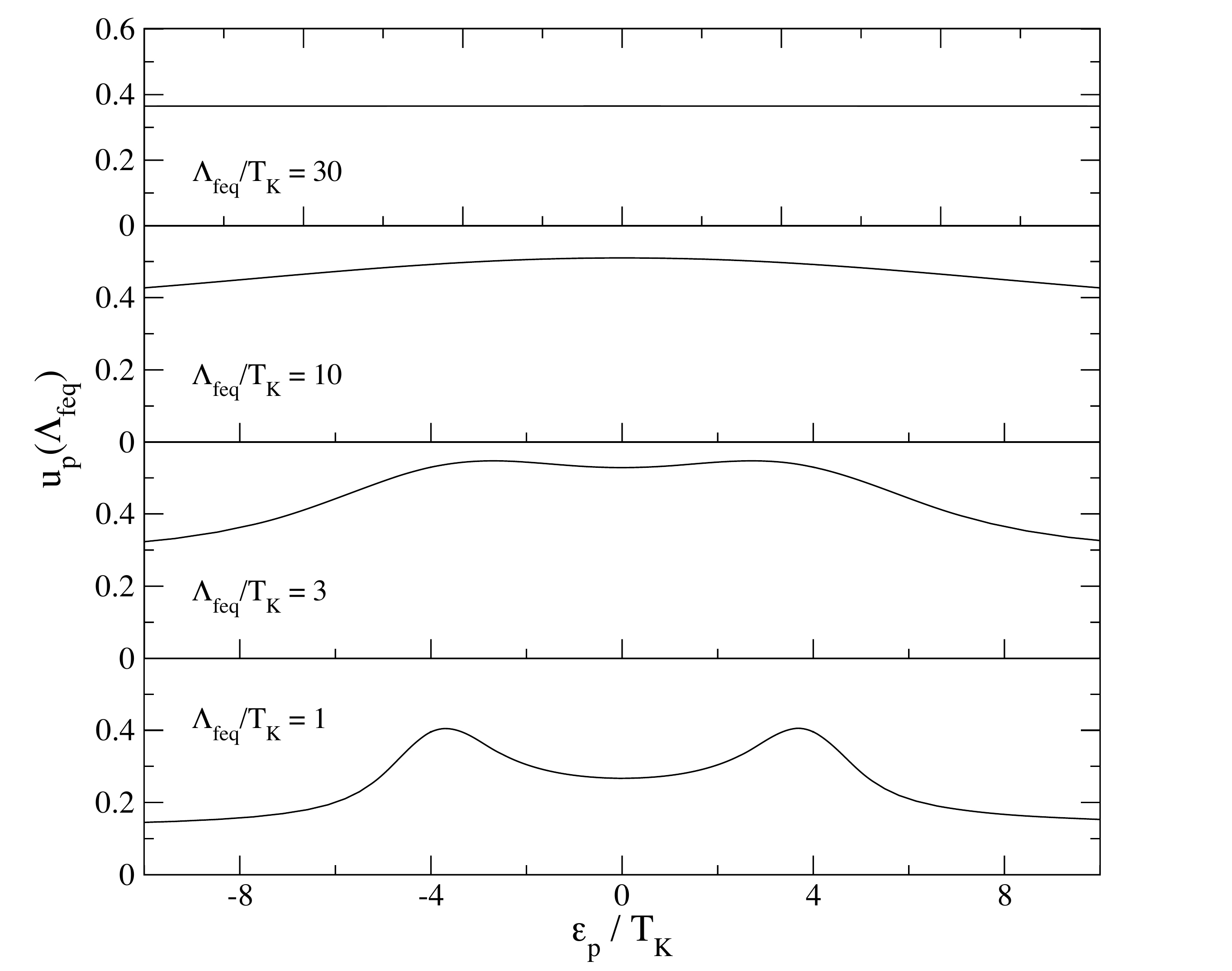}
\caption{Running coupling constants $u_p(\Lambda_{\rm feq})$ at various
points during the flow for a symmetric Kondo dot with $V/T_{\rm K}=8.0$.
One observes the
initial buildup of strong-coupling behavior until
$\Lambda_{\rm feq}/T_{\rm K}\approx 3$, followed by decreasing coupling
constants even at the left and right Fermi surfaces due to decoherence:
there is no 2-channel strong-coupling divergence.}
\label{fig_flowup_sym}    
\end{figure}

\begin{figure}[t]
\includegraphics[clip=true,width=8.0cm]{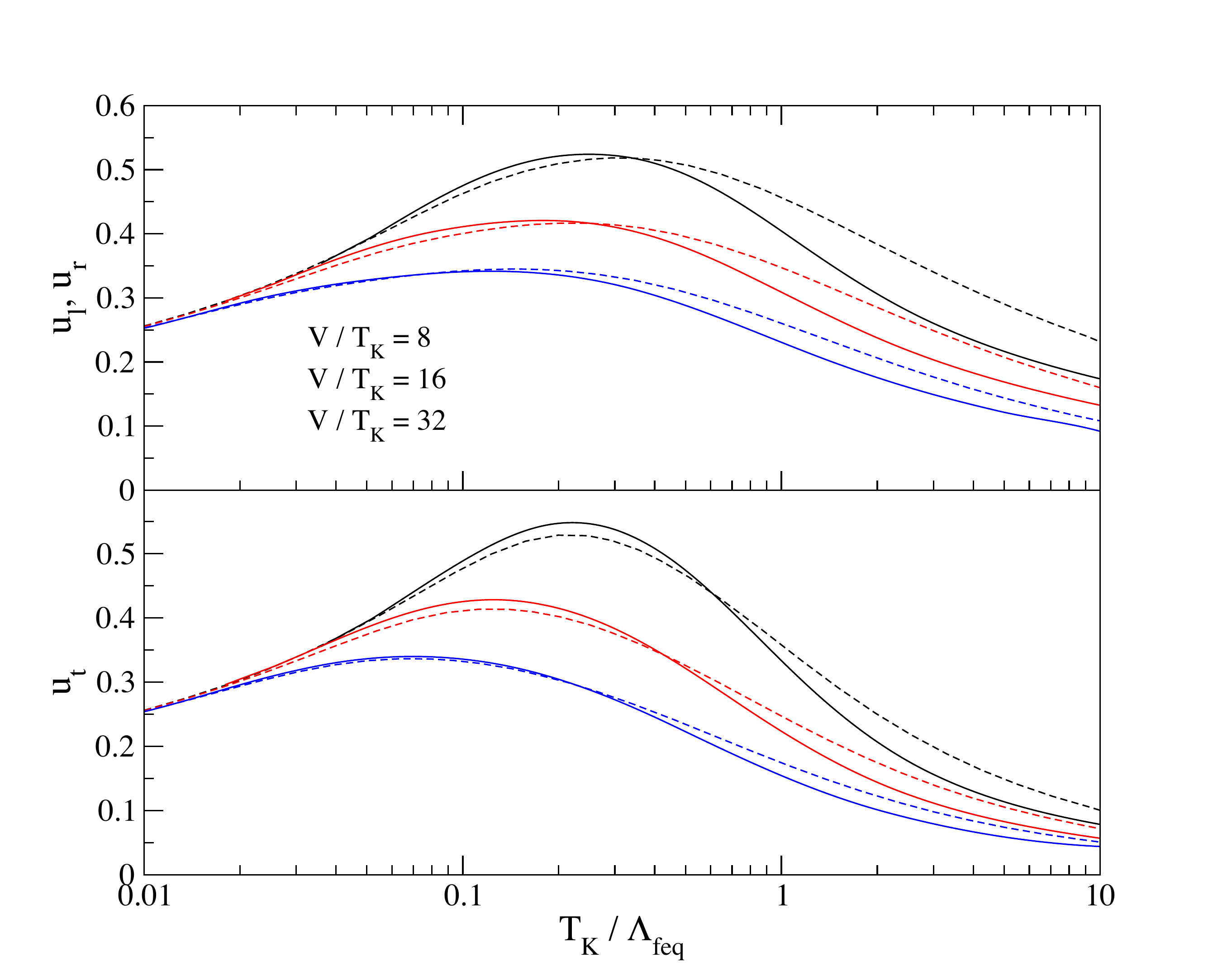}
\caption{Comparison of the flow of the coupling constants $u_l$ and $u_t$
from the full numerical solution of (\ref{neq_J2loopa}) and (\ref{neq_K2loopa}) 
with the effective equations (\ref{eq_ut}) and (\ref{eq_ul}).
Depicted here is a symmetric Kondo dot (therefore $u_l=u_r$).
The full lines refer to the full numerical solution and the dashed
lines to the effective equations. Results are shown for various
ratios $V/T_{\rm K}$ labelling the pairs of curves from top to bottom.}
\label{fig_comparison_sym}    
\end{figure}

\begin{figure}[t] 
\includegraphics[clip=true,width=8.0cm]{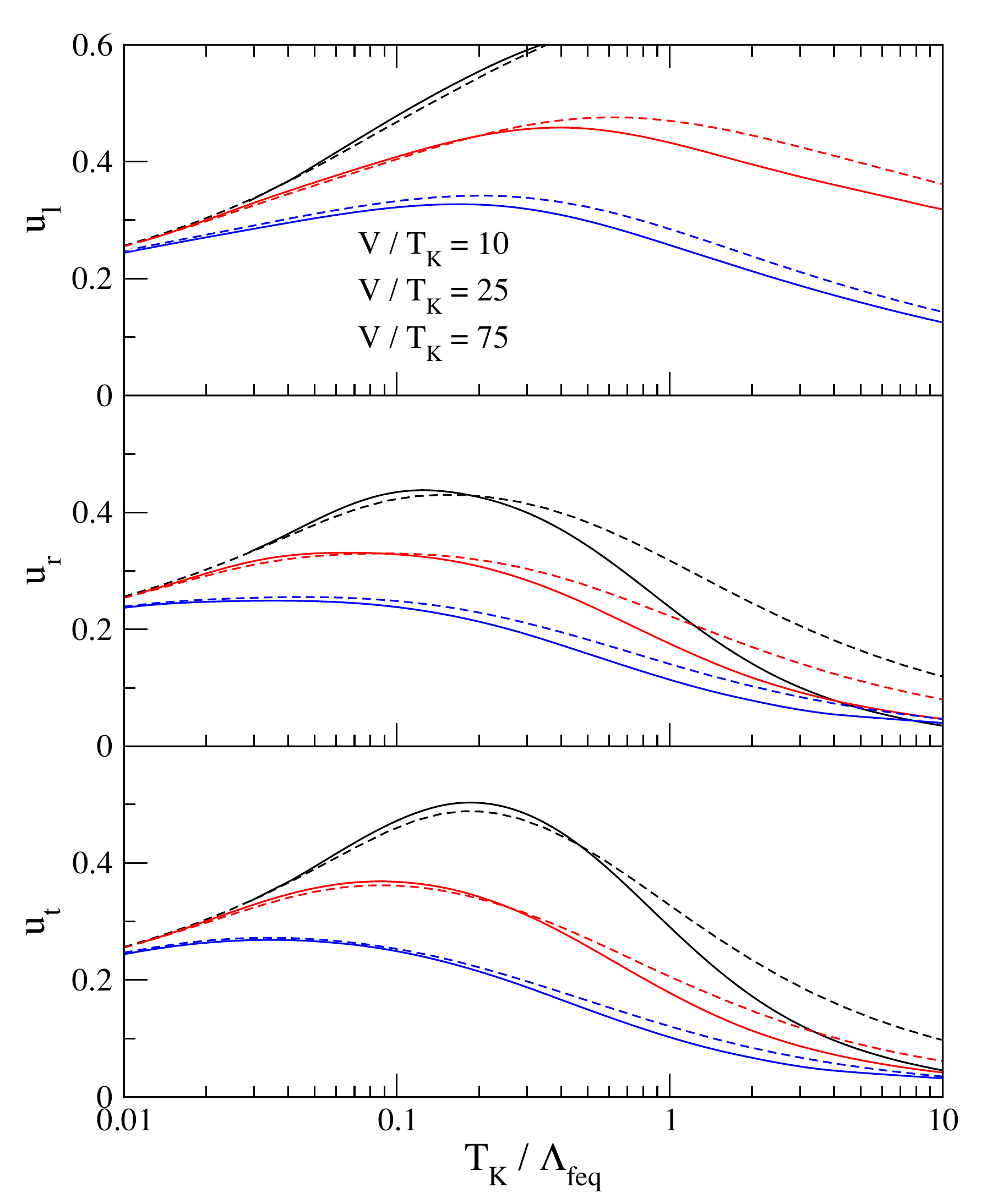}
\caption{
Same as Fig.~\protect\ref{fig_comparison_sym}
for the flow of $u_l$, $u_r$ and $u_t$ for an
asymmetric Kondo dot with $r=\Gamma_l/\Gamma_r=2$.
The curves for $V/T_{\rm K}=10$ should only be considered 
until $T_{\rm K}/\Lambda_{\rm feq}\approx 0.3$ since the
scaling flow leaves the weak-coupling regime for smaller
flow parameters~$\Lambda_{\rm feq}$.}
\label{fig_comparison_asym}    
\end{figure}

Figs.~\ref{fig_comparison_sym} and \ref{fig_comparison_asym}
shows a comparison of the flow of the
coupling constants $u_l(\Lambda_{\rm feq}), u_r(\Lambda_{\rm feq})$
and $u_t(\Lambda_{\rm feq})$ from the full numerical solution
with the solution of (\ref{eq_ut}), (\ref{eq_ul}) and (\ref{eq_ur}).
One can see that our approximations in going from $N\times N$
differential equations to just 3~differential equations were 
very accurate: the resp.~solutions agree very well. 
This agreement is somehow obvious in the initial phase of
the flow where the voltage bias plays no role. Therefore
one can safely start the full numerical solution at 
$\Lambda_{\rm feq}= 4V$ and not waste computer time with larger
values of the initial cutoff. The curves agree very well until
one reaches the decoherence scale (maximum of the coupling constants),
which was important for determining the phase diagram. The small
absolute differences at the decoherence scale 
between the full numerical solution and 
the analytical approximation get somehow amplified if one looks at
the relative error
in the later phase of the flow. This will, however, anyway turn out to
be unimportant for the evaluation of observables (which are essentially 
determined by the absolute values of the coupling constants
at the decoherence scale).
In general one can notice
from Figs.~\ref{fig_comparison_sym} and \ref{fig_comparison_asym} 
that  the accuracy of our
analytical approximations (\ref{eq_ut}), (\ref{eq_ul}) and (\ref{eq_ur})
becomes increasingly better if the coupling constants remain small during the 
flow, that is for larger values of the voltage bias $V/T_{\rm K}$. 
These obsersations
therefore justify a posteriori our previous analytical analysis based on
(\ref{eq_ut}), (\ref{eq_ul}) and (\ref{eq_ur}).

\section{Observables}
\label{ch_observables}
\subsection{Observables in the Flow Equation Framework}
\label{chneqobservables}
The evaluation of dynamical quantities like correlation functions within
the flow equation framework is rather different from conventional
many-body techniques. The key observation is that while the time evolution
becomes trivial in the diagonal basis for $B\rightarrow\infty$, the
structure of observables can be very complicated in this basis.\cite{Kehreinbook}
In order to use the time evolution with respect to the diagonal Hamiltonian $H(B=\infty)$,
one first needs to transform an observable~$O$ with the same unitary
transformations as the Hamiltonian itself (\ref{eq_feq})
\beq
\frac{dO(B)}{dB}=[\eta(B),O(B)] \ ,
\label{eqtrfO}
\eeq
where $O(B=0)=O$ is the operator in the original basis. 
Once has found $O(B=\infty)$, it becomes easy to work out its equilibrium correlation
functions: We are for example interested in
\bea
C_{\rm eq}(t)&=&\langle {\rm GS}|\, O(t)\,O(0) \,|{\rm GS}\rangle \nn \\
&=&\langle {\rm GS}|\,e^{iHt}\,O\, e^{-iHt}\,O\,|{\rm GS}\rangle \ ,
\label{def_Ct}
\eea
where $|{\rm GS}\rangle$ is the interacting ground state of the system. We denote
the unitary transformation that connects the initial with the final basis with
$U(B=\infty)$, therefore 
\bea
\tilde H=H(B=\infty)&=&U(B=\infty)\,H(B=0)\,U^\dagger(B=\infty) \nonumber \\
\tilde O=O(B=\infty)&=&U(B=\infty)\,O(B=0)\,U^\dagger(B=\infty) \nonumber
\eea
where we have introduced the notation~$\tilde{}$ that designates an operator in
the transformed basis. One can (at least formally) construct $U(B=\infty)$ from 
(\ref{eq_feq}) in the following way:
\beq
U(B)\stackrel{\rm def}{=}T_B \,\exp\left(\int_0^B dB'\,\eta(B')\right) \ .
\eeq
Here $T_B$ denotes $B$-ordering in the same way as the usual time-ordering:
the generator $\eta(B'_i)$ with the largest~$B'_i$ is commuted all the way
to the left, next comes the second largest, etc. Next one inserts identities
$U^\dagger(B=\infty)\,U(B=\infty)=1$ everywhere in (\ref{def_Ct}) and 
arrives at
\beq
C_{\rm eq}(t)=\langle {\rm GS}|\,U^\dagger(B=\infty)\,e^{i\tilde Ht}\,\tilde O\,
e^{-i\tilde Ht}\, \tilde O\,U(B=\infty)\,|{\rm GS}\rangle \ .
\label{unitaryCt}
\eeq
A key observation is that $U(B=\infty)\,|{\rm GS}\rangle$ is just the
ground state of the diagonal Hamiltonian~$\tilde H$, which is therefore trivial
to write down (depending on the system for example just the vacuum, the 
noninteracting Fermi sea, etc.)
\beq
|{\rm GS}_{\tilde H}\rangle=U(B=\infty)\,|{\rm GS}\rangle \ .
\label{GStildeH}
\eeq 
The evaluation of equilibrium correlation functions therefore reduces to
\beq  
C_{\rm eq}(t)=\langle {\rm GS}_{\tilde H}|\,e^{i\tilde Ht}\,\tilde O\,
e^{-i\tilde Ht}\, \tilde O\,|{\rm GS}_{\tilde H}\rangle \ .
\label{Ceqt}
\eeq
Since it is usually straighforward to solve the Heisenberg equations of
motion for an operator with respect to the diagonal Hamiltonian~$\tilde H$, this
equation has been used as the starting point for the evaluation of correlation 
functions in various models like the spin-boson model\cite{Kehrein96,Kleff},
the Kondo model\cite{Kehrein_KM}, etc. 
Before proceeding along the same lines we first need to reconsider the
changes in this derivation for a non--equilibrium model like the Kondo model
with voltage bias.

The fundamental difference from the equilibrium situation is that now
we are {\em not\/} interested in evaluating observables in the
ground state. Rather, the relevant question is
the behavior of observables in the {\em steady state\/} that evolves after 
switching on the coupling between the left and right lead and then waiting
long enough. This is also the key idea of the Keldysh approach
and in this manner one circumvents 
the generally unanswered fundamental question on how to construct a non--equilibrium
steady state. We follow the same
idea here by replacing (\ref{def_Ct}) with its non--equilibrium generalization
\beq
C_{\rm neq}(t)\stackrel{\rm def}{=}\lim_{t_w\rightarrow\infty} C(t,t_w) \ ,
\label{def_Cneqt}
\eeq
where
\bea
\lefteqn{
C(t,t_w)\stackrel{\rm def}{=}
\langle \Psi_i|\,O(t+t_w)\,O(t_w)\,|\Psi_i\rangle } \\
&=&\langle \Psi_i|\,e^{iH(t+t_w)}\,O\, e^{-iH(t+t_w)}\,e^{iHt_w}\,O\,e^{-iHt_w}\,|\Psi_i\rangle \ .
\nonumber
\eea
Here we assume that the thermodynamic limit is always taken before 
sending the waiting time~$t_w$ for the measurement to infinity in (\ref{def_Cneqt}).
Notice that in general $C_{\rm neq}(t)$ could depend on the initial state~$|\Psi_i\rangle$
and in the sequel we will use an initial state that is
directly related to the experimental situation: $|\Psi_i\rangle$ is the ground state
of the system when the two leads are not coupled via the Kondo impurity, that is
the noninteracting Fermi sea in both the left and right lead separately.

We now use the same unitary transformations as going from (\ref{def_Ct})
to (\ref{unitaryCt})
\bea
\lefteqn{C(t,t_w)=} \label{Cttw} \\
&=&\langle \Psi_i|\,U^\dagger(B=\infty)\,e^{i\tilde H(t+t_w)}\,\tilde O\, 
e^{-i\tilde H(t+t_w)} \,U(B=\infty)  \nonumber \\
&&\qquad \times \,U^\dagger(B=\infty)
\,e^{i\tilde Ht_w}\,\tilde O\,e^{-i\tilde Ht_w}\,U(B=\infty)\,|\Psi_i\rangle \ . 
\nonumber 
\eea
The additional unitary transformation of the time-evolved operator in the diagonal
basis just maps back to the original basis, where the initial state is specified.
This {\em forward-backward scheme\/} has also been employed in other models \cite{HacklKehrein} 
where one is interested in real time evolution from some given initial state.
The backward transformation part of this scheme can be implemented by
solving (\ref{eqtrfO}) from $B=\infty$ to $B=0$: the initial condition at
$B=\infty$ is just the time-evolved operator in the diagonal basis. In order to evaluate
(\ref{def_Cneqt}) from this expression we next need to say something about the explicit transformation
behavior of the spin operator in the next subsection.

\subsection{Transformation of the Spin Operator}
The transformation of the spin operator $\vec S$ under the flow equation transformation 
has already been worked out in Ref.\cite{Kehreinbook} and we only give the key steps here.
One makes the following ansatz for the flowing observable (\ref{eqtrfO}):
\beq
\vec S(B)=h(B)\,\vec S+i\sum_{t',t}\gamma_{p'p}(B)\,\vec S\times \vec s_{p'p}+O(u^{2})
\label{eqtrfS}
\eeq 
and derives the flow equations for the coefficients from (\ref{eqtrfO}):
\bea
\frac{dh}{dB}&=&\sum_{p',p}(\eps_{p'}-\eps_{p})\,J_{p'p}\gamma_{pp'}\,n_{f}(p')\left(1-n_{f}(p)\right) \nn \\
\frac{d\gamma_{p'p}}{dB}&=&h\,(\eps_{p'}-\eps_{p})\,J_{p'p} - \frac{1}{4} \sum_{u} \big( (\eps_{p'}-\eps_{u}) J_{p'u}\gamma_{up} \nn\\
& \ & + (\eps_{p}-\eps_{u}) J_{up}\gamma_{p'u} \big)
(1-2n_{f}(u)) \ .
\label{feq_hgamma}
\eea
Initially $h(B=0)=1$, $\gamma_{p'p}(V=0)=0$ and under the flow 
one always finds $h(B=\infty)=0$ (see Refs.\cite{Kehrein96,Kehreinbook}).
One easily identifies this scale where $h(B)$ rapidly drops to~0 by noticing that
$\gamma_{p'p}(B)$ is initially only generated in order~$u$, therefore $h(B)$ initially remains
nearly constant since its differential equation only generates terms in order~$u^{2}$.
This picture breaks down on the scale $\Lambda_{\rm feq}\sim \Gamma_{\rm rel}$
where $h(B)$ starts to decay algebraically~$\propto B^{-1/4}$:
\begin{itemize}
\item[i)] Nonequilibrium (voltage bias~$V$):
\beq
\Gamma_{\rm rel}(V)  =  \sqrt{2\pi}u^2(\Lambda_{\rm feq}=V) \frac{V}{(1+r)(1+r^{-1})}
\label{eqdefGammaV}
\eeq
\item[ii)] Equilibrium (nonzero temperature):
\beq
\Gamma_{\rm rel}(T)= \sqrt{2\pi}u^2(\Lambda_{\rm feq}=T)\ T
\label{eqdefGammaT}
\eeq
\end{itemize}
Notice that these energy scales are identical with the flow scale (\ref{def_Gammarel})
where the transmutation of the third order terms in the coupling constant
into linear terms occurs. At this point it is easy to understand this observation:
When $h(B)$ becomes small, the spin operator (\ref{eqtrfS}) completely decays into
its entangled form $\vec S\times\vec s_{p'p}$. In the flow of the Hamiltonian this
amounts to transforming the marginally relevant interaction term $J_{t't}\,\vec S\cdot s_{t't}$
into the irrelevant $K$-term, $K_{t't,p'p}\,\vec S\cdot (\vec s_{t't}\times \vec s_{p'p})$.
This implies that the flow equations for $u_{t}, u_{l}, u_{r}$ (\ref{eq_utb}--\ref{eq_urb})
become weak-coupling for $\Lambda_{\rm feq}\lesssim \Gamma_{\rm rel}$ as discussed
before. In the next section we will identify the energy scale where the decoherence effects
become important in the Hamiltonian flow with the width of the zero frequency peak of the
correlation function, that is the physical relaxation rate.

\subsection{Dynamical Spin Response and Correlation Functions}
\label{ch_dynamicspin}
In the following we will use the shorthand $\tilde{\gamma}_{p'p}$ for $\gamma_{p'p}(B=\infty)$.
From (\ref{Ceqt}) one derives the following expression for the 
symmetrized spin--spin correlation function in equilibrium at nonzero temperature:
\bea
C_{\rm eq}(t)&\stackrel{\rm def}{=}&\frac{1}{2}\,\langle \{S_{z}(0),S_{z}(t)\}\rangle \nn \\
&=&-\frac{1}{2}\sum_{p',p}\sum_{q',q}\tilde{\gamma}_{p'p}\tilde{\gamma}_{q'q}\,
e^{i(\eps_{q'}-\eps_{q})t}\, \nn \\
&&\times 
\langle {\rm FS}|\{(\vec S\times \vec s_{p'p})^{z},(\vec S\times \vec s_{q'q})^{z}\}|{\rm FS}\rangle \nn \\
&=& \frac{1}{8}\sum_{p',p}\tilde{\gamma}_{p'p}^{2}\,
e^{i(\eps_{p'}-\eps_{p})t}  \left(n_{f}(p)\left(1-n_{f}(p')\right)\right.\nn\\
&&\left.+ n_{f}(p')\left(1-n_{f}(p)\right)\right) 
\label{spinspineq}
\eea
Here $|{\rm FS}\rangle$ is the noninteracting Fermi sea and $n_f(p)$ the finite
temperature Fermi-Dirac distribution.

For the symmetrized spin--spin correlation function in the non--equilibrium 
steady state (\ref{def_Cneqt}) we need to perform the backtransformation (\ref{Cttw})
of $\tilde S_z(t_w)$ for large~$t_w$. Due to the phases $\exp[i(\eps_{q'}-\eps_{q})\,t_w]$
that the matrix elements $\tilde{\gamma}_{q'q}$ acquire, one can easily see that 
integrating (\ref{feq_hgamma}) back from $B=\infty$ to $B=0$ only leads to higher order
contributions in the flow equations. For large times we can therefore say
\beq
S_z(t_w)=\tilde S_z(t_w)\:\left(1+O(g)\right) \ .
\eeq
This implies for the spin--spin correlation function in the steady state (\ref{def_Cneqt})
\bea
C_{\rm neq}(t)&=&\frac{1}{8}\sum_{p',p}\tilde{\gamma}_{p'p}^{2}\,
e^{i(\eps_{p'}-\eps_{p})t}  \left(n_{f}(p)\left(1-n_{f}(p')\right)\right.\nn\\
&&\left.+ n_{f}(p')\left(1-n_{f}(p)\right)\right) 
\label{spinspinneq}
\eea
plus higher order corrections in the running coupling constant.
Here we now have to insert the occupation numbers (\ref{occnonequ}) for the problem with voltage bias.

Both the nonzero temperature equilibrium result (\ref{spinspineq})
and the steady state non--equilibrium result (\ref{spinspinneq}) have the same structure
and can therefore be analyzed together:
Fourier transformation yields 
\bea
C(\omega)&=&\int dt \,e^{i\omega t}\,C(t) \nn \\
&=&\frac{\pi}{4}\sum_{p}\tilde{\gamma}_{p+\omega,p}^{2} \left(n_{f}(p)\left(1-n_{f}(p+\omega)\right) \right.\nn\\
&&\left.+ n_{f}(p+\omega)\left(1-n_{f}(p)\right)\right) \label{corr_func}
\eea
where the notation $p+\omega$ stands for the state with energy $\eps_{p}+\omega$.
Due to the perturbative nature of the transformation the sum rule 
\beq
\int_{-\infty}^{\infty} d\omega\, C(\omega)=2\pi\,\langle S_{z}^{2}\rangle =\frac{\pi}{2}
\eeq
is not fulfilled exactly. For high voltage bias or high temperature this violation is typically only of order one percent,
which becomes up to ten percent for voltage bias or temperature of order~$T_K$. 
 
The same calculation for the spin susceptibility
\beq
\chi(t)\stackrel{\rm def}{=}-i\,\Theta(t)\,\langle [S_{z}(t),S_{z}(0)] \rangle 
\eeq
yields the imaginary part of its Fourier transform
\bea
\chi''(\omega)&=&\frac{\pi}{4}\sum_{p}\tilde{\gamma}_{p+\omega,p}^{2} \nn 
\left(n_{f}(p)\left(1-n_{f}(p+\omega)\right) \right.\nn\\
&& \left.- n_{f}(p+\omega)\left(1-n_{f}(p)\right)\right)\ .\label{response_func}
\eea
The real part follows via a Kramers--Kronig relation, in particular the static spin susceptibility~$\chi_{0}$
is given by
\beq
\chi_{0}=\frac{1}{\pi}\,\int_{-\infty}^{\infty} d\omega\:
\frac{\chi''(\omega)}{\omega} \ .
\label{defchi0}
\eeq
\begin{figure}[t]
\includegraphics[clip=true]{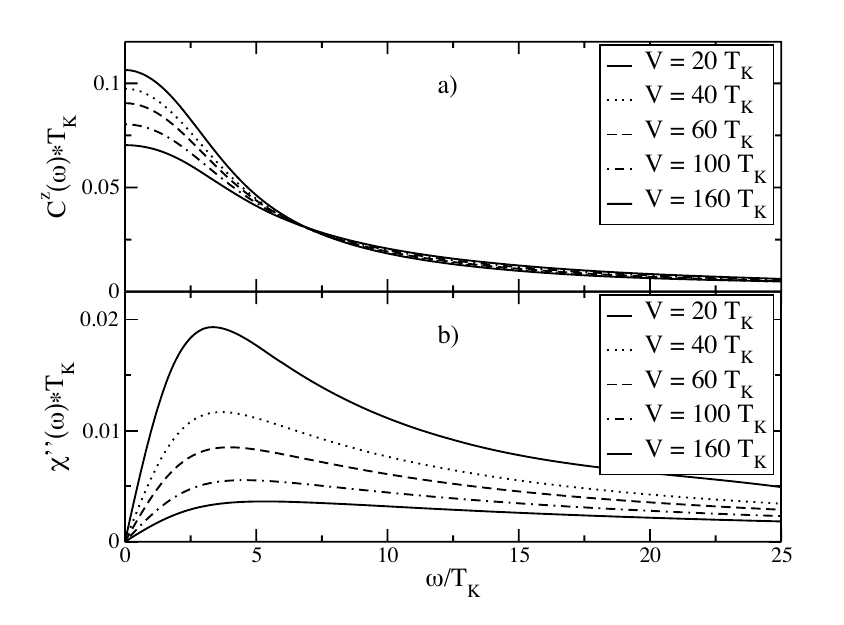}
\caption{Universal curves for the spin-spin correlation function $C(\omega)$ and the imaginary part of the dynamical spin
susceptibility $\chi''(\omega)$ in non-equilibrium (symmetric coupling $r=1$).}
\label{Fig_corr_V}    
\end{figure} 
\begin{figure}[t]
\includegraphics[clip=true]{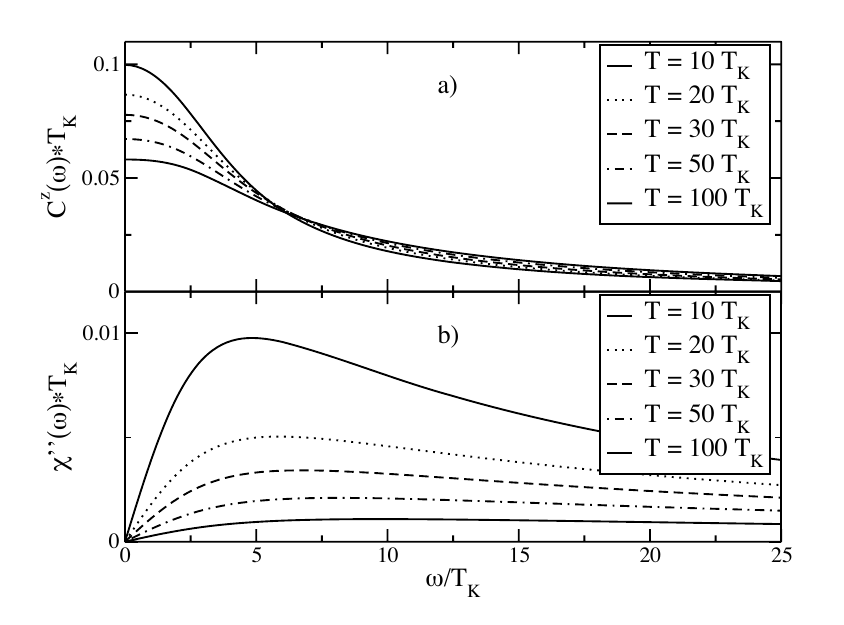}
\caption{Universal curves for the spin-spin correlation function $C(\omega)$ and the imaginary part of the dynamical spin
susceptibility $\chi''(\omega)$ in equilibrium for various temperatures.}
\label{Fig_corr_T}    
\end{figure}
Notice that the fluctuation--dissipation theorem\cite{FlucDiss}
\beq
\chi''(\omega) = \tanh\left( \frac{\omega}{2T} \right) C(\omega) \label{flucdiss_def}
\eeq
is fulfilled exactly in equilibrium. In non-equilibrium the fluctuation-dissipation relation is fulfilled
approximately in the zero frequency limit if one inserts an effective temperature motivated by the
comparison of (\ref{eqdefGammaV}) and (\ref{eqdefGammaT}):
\beq
T_{\rm eff} = \frac{V}{(1+r)(1+r^{-1})}\ .
\label{eqdefTeff}
\eeq

Typical curves for the spin-spin correlation function and the imaginary part of the dynamical spin susceptibility are shown in Fig. \ref{Fig_corr_V} for
non-equilibrium (for varying voltage bias) and in Fig. \ref{Fig_corr_T} for equilibrium (for varying temperature).
The spin-spin correlation function is a symmetric function of $\omega$ with a zero frequency peak of width $\Gamma_{\rm rel}$. The
imaginary part of the
dynamical spin susceptibility is an antisymmetric function of $\omega$ and has its maximum at
$\chi''(\omega\approx \Gamma_{\rm rel})$.
An approximate analytical solution of (\ref{feq_hgamma}) yields\cite{Kehreinbook}
(see also Ref.~\cite{Rosch04b}):
\beq
C(\omega) \sim \left\{
\begin{array}{ll}
\frac{\ds 1}{\ds \Gamma_{\rm rel}} & \text{~~for }|\omega|\lesssim \Gamma_{\rm rel} \\
\ & \ \\
\frac{\ds\Gamma_{\rm rel}}{\ds\omega^2} & \text{~~for } \Gamma_{\rm rel} \lesssim |\omega|\lesssim T_{\rm eff}\\
\ & \ \\
\frac{\ds u^2(\Lambda_{\rm feq}=|\omega|)}{\ds |\omega|} & \text{~~for } T _{\rm eff} \lesssim |\omega|
\end{array}
\right. \label{corr_detail}
\eeq 
\beq
\chi''(\omega) \sim \left\{
\begin{array}{ll}
\frac{\ds u^2(\Lambda_{\rm feq}=T,V)}{\ds \Gamma_{\rm rel}^2}\,\omega & \text{~~for }|\omega|\lesssim \Gamma_{\rm rel} \\
\ & \ \\
\frac{\ds u^2(\Lambda_{\rm feq}=|\omega|)}{\ds \omega} & \text{~~for } \Gamma_{\rm rel} \lesssim |\omega|\ .
\end{array}
\right. \label{chi2_detail}
\eeq 

Notice that the zero frequency peak of the correlation function
is directly related to the flow scale where $h(B)$ in the transformation of the spin operator
starts to deviate noticeably from~1 since essentially all the spectral weight is contained
in the energy interval $O(\Gamma_{\rm rel})$ according to (\ref{corr_detail}). This provides
the desired physical interpretation of~$\Gamma_{\rm rel}$ defined in (\ref{def_Gammarel}):
The energy scale where the Kondo coupling becomes irrelevant can be identified with the 
spin relaxation rate.

\begin{figure}[t]
\includegraphics[clip=true]{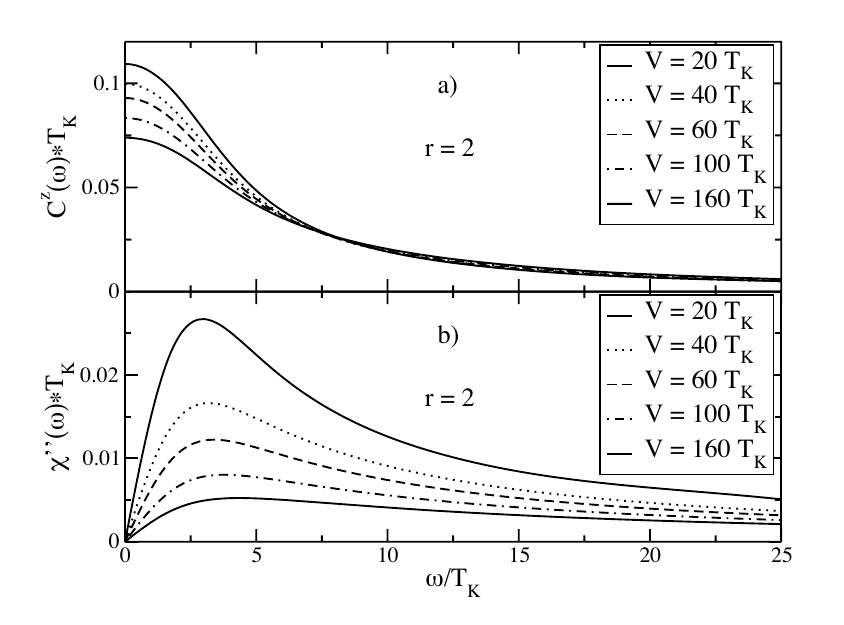}
\caption{Universal curves for the spin-spin correlation function $C(\omega)$ and the imaginary part of the dynamical spin susceptibility
$\chi''(\omega)$ in non-equilibrium for asymmetric coupling of the leads $r=2$.}
\label{Fig_corr_R_2}    
\end{figure}
\begin{figure}[t]
\includegraphics[clip=true]{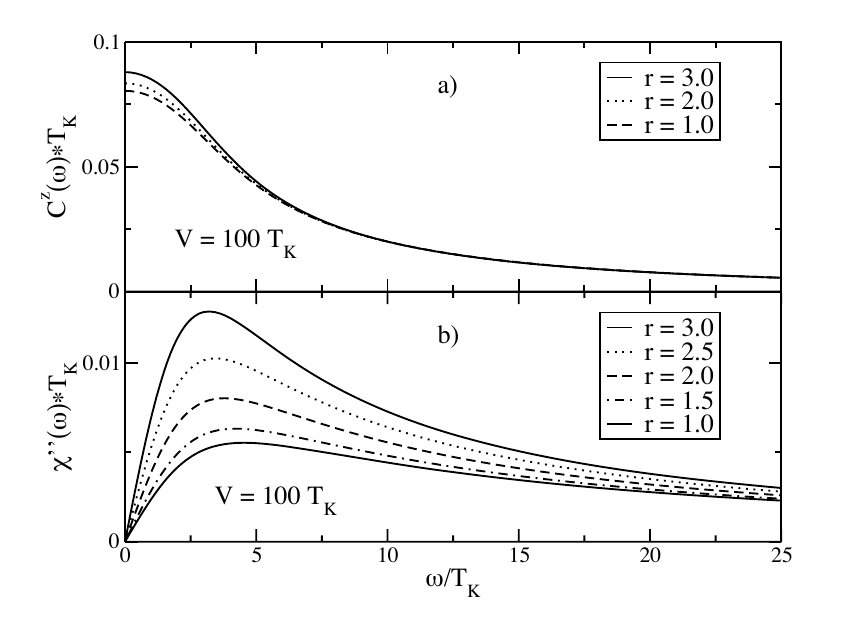}
\caption{Universal curves for the spin-spin correlation function $C(\omega)$ and the imaginary part of the dynamical spin susceptibility
$\chi''(\omega)$ in non-equilibrium for asymmetric coupling of the leads. The voltage bias is held fixed at $V = 100\ T_K$.}
\label{Fig_corr_R_V_100_T_K}    
\end{figure}
In Fig.~\ref{Fig_corr_R_2} we show the spin-spin correlation function and the imaginary part of the 
dynamical spin susceptibility for asymmetric coupling to the leads with $r=2$.
In Fig.~\ref{Fig_corr_R_V_100_T_K} we depict the dependence on the asymmetry paramter $r$ at fixed voltage bias.
Increasing the asymmetry parameter~$r$ leads to a smaller decoherence rate since the effective temperature
(\ref{eqdefTeff}) becomes smaller.
In agreement with (\ref{corr_detail}) and (\ref{chi2_detail}) one therefore notices that mainly the low frequency properties are affected
with the maxima of the
curves increasing for larger asymmetry.

\subsection{Static Spin Susceptibility}
One of the most celebrated results of Kondo physics is the behavior of the static spin susceptibility~$\chi_{0}$
as a function of temperature. This behavior encompasses the screening of the impurity spin via the
formation of a Kondo bound state upon lowering temperature. While it is easy to deduce the 
reduction of the magnetic moment with scaling techniques,
\beq
\chi_{0}(T)=\frac{1}{4T} \left( 1-\frac{1}{\ln(T/T_K)}\right) \ ,
\label{eqchi0logcorr}
\eeq
the correct result of a finite nonzero spin susceptiblity at zero temperature, $\chi_{0}(T=0)$,
was first obtained in the seminal application of the numerical renormalization group by Wilson \cite{Wilson_NRG}.
Later the Bethe ansatz \cite{BA_thermo,BA_scales}  gave the complete analytical solution leading to the high 
temperature expansion
\bea
\chi_{0}(T) & = & \frac{1}{4T} \left( 1-\frac{1}{\ln(T/T_K)} \right.\label{chi0_BA}\\
& \ & \left. -\frac{\ln(\ln(T/T_K))}{2(\ln(T/T_K))^2} +
{\cal O}\left((\ln(T/T_K))^{-2} \right)\nn
\right)\ .
\label{eqBAchi0}
\eea
{\em  Likewise, the behavior of the static spin susceptibility as a function of voltage
bias is of fundamental importance for understanding non-equilibrium Kondo physics.} In the sequel
we will focus on the case of zero temperature, zero magnetic field and nonzero voltage bias.
The case of nonzero magnetic field will be discussed in a separate publication. The leading order
result for $\chi_{0}(T,V)$ for large temperature or large voltage bias was worked out by 
Parcollet and Hooley \cite{Parcollet02}:
\beq
 \chi_{0}(T=0,V)=\frac{(1+r)(1+r^{-1})}{4V}=\frac{1}{4T_{\rm eff}} \ .
\label{chi0ParHol}
 \eeq
This naturally raises the question about the logarithmic corrections to this behavior, similar
to the key result (\ref{eqchi0logcorr}) of equilibrium Kondo physics. This question has to date
only been answered for the leading $\ln^2 (V/T_{\rm K})$-corrections, which, however, vanish
if the Kondo model can be derived from an underlying Anderson impurity model (see Sect.~II.A).
The logarithmic terms proportional to $\ln (V/T_{\rm K})$ have not yet been calculated 
completely, which means that the current status of the non--equilibrium calculation would
miss the finite temperature logarithmic correction in equilibrium (\ref{eqchi0logcorr}).
More details about these calculations and results can be found in Refs.~\cite{Rosch03,Rosch04a,Rosch05}
based on perturbative RG and Keldysh techniques.

\begin{figure}[t]
\includegraphics[clip=true]{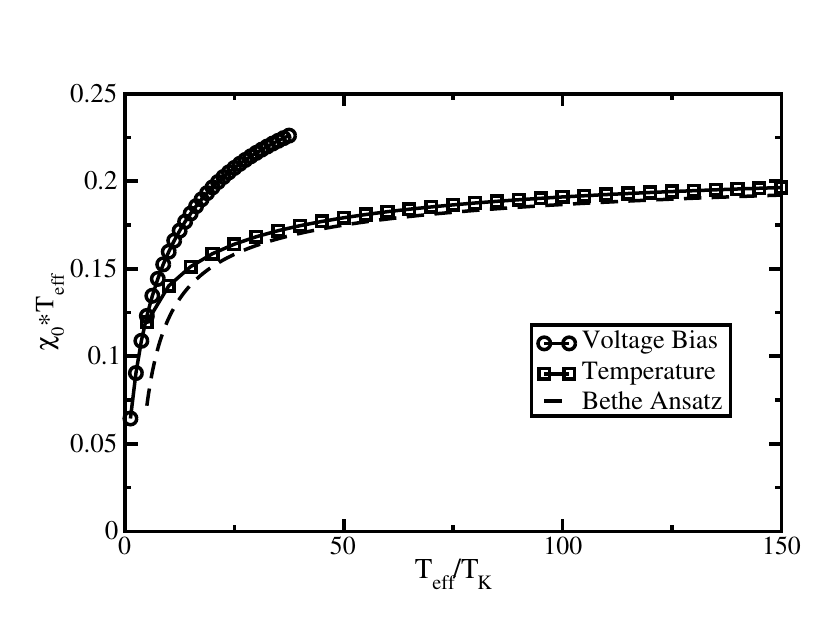}
\caption{Flow equation results for the static spin susceptibility $\chi_{0}$ for nonzero temperature
(equilibrium) and for nonzero voltage bias (non-equilibrium, symmetric coupling $r=1$). The data
is plotted as a function of the effective temperature $T_{\text{eff}}=V/4$ in the case of voltage bias 
(while $T_{\text{eff}}\stackrel{\rm def}{=}T$ for nonzero temperature). For comparison we
show the leading order Bethe ansatz result (\ref{eqBAchi0}) for the equilibrium susceptibility, 
which agrees very well with the finite temperature flow equation data.}
\label{Fig_chi0}    
\end{figure}

\begin{figure}[t]
\includegraphics[clip=true,width=9cm]{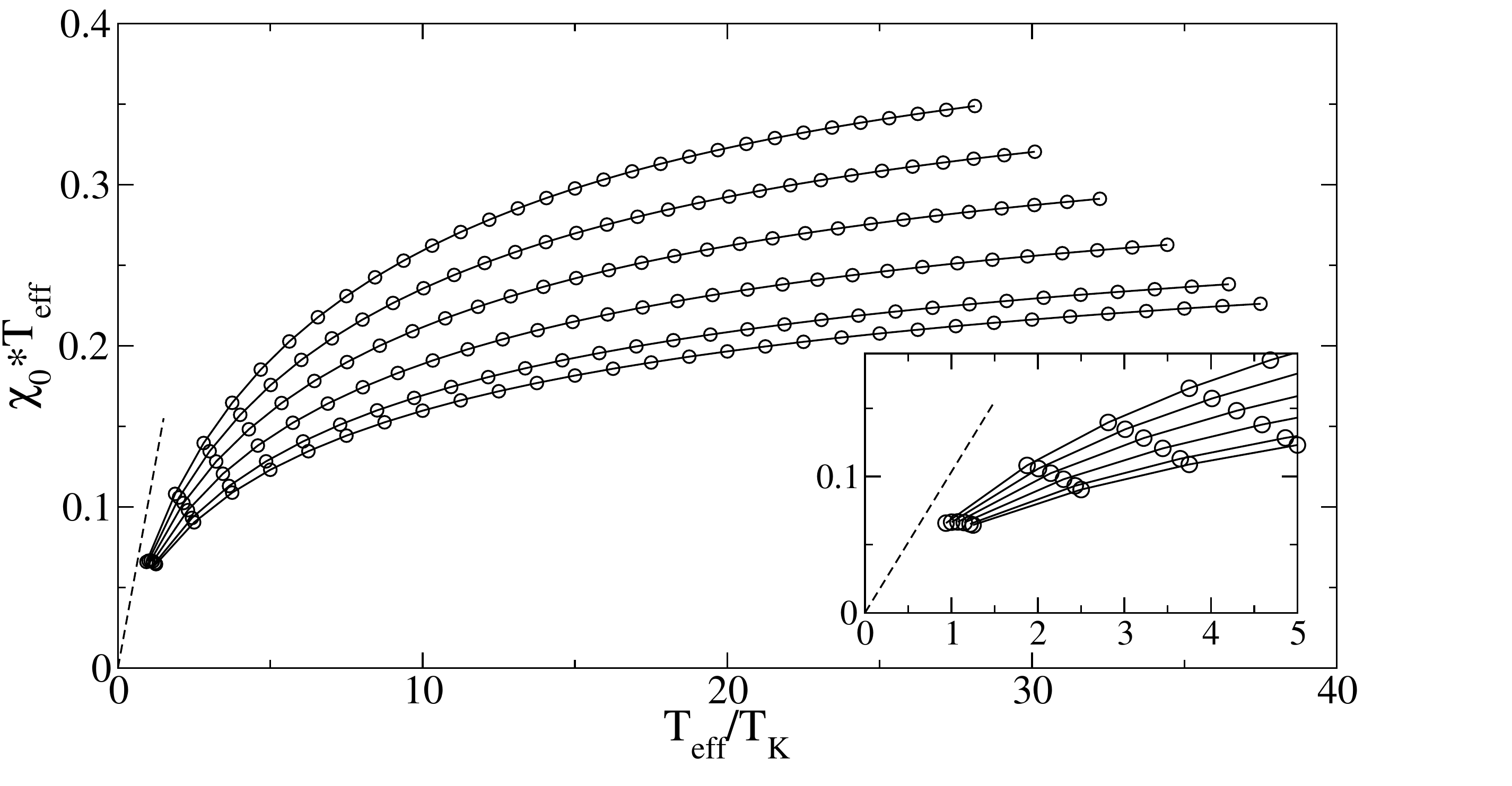}
\caption{Flow equation results for the static spin susceptibility $\chi_{0}$ for nonzero voltage bias 
for various asymmetry parameters~$r$. The asymmetry parameter increases from bottom to top:
$r=1.0, 1.4, 1.8, 2.2, 2.6, 3.0$. 
The data is plotted as a function of $T_{\text{eff}}=V/(1+r)(1+r^{-1})$, Eq.~(\ref{eqdefTeff}).
The dashed line is an exact result for the behavior in the $T_{\rm eff}\rightarrow 0$ limit
independent of~$r$, see text.
Lines are guide to the eye only and the numerical error of the datapoints is at most~10\%.
The inset shows a blowup of the small voltage bias region.}
\label{Figchi0RTeff}    
\end{figure}

Within the flow equation framework it is straightforward to derive the leading term
(\ref{chi0ParHol}) {\em analytically\/}, but the corrections to it can only be worked
out from the full numerical solution of the system of differential equations derived
in the previous chapters. Numerical results for the static spin susceptibility obtained
in this way via (\ref{defchi0}) are shown in Figs.~\ref{Fig_chi0} and~\ref{Figchi0RTeff}.

The comparison (Fig.~\ref{Fig_chi0}) of the Bethe ansatz result (\ref{eqBAchi0}) with the flow equation
data points in equilibrium as a function of temperature shows very good
agreement for large $V/T_{\rm K}$. This demonstrates that the flow equation calculation
presented here contains the leading logarithmic corrections and can address this issue
also in non--equilibrium. 

Flow equation results for this non--equilibrium spin susceptibility are plotted
in Fig.~\ref{Fig_chi0} for symmetric coupling to the leads ($r=1$) as a function
of the effective temperature $T_{\rm eff}=V/4$, Eq.~(\ref{eqdefTeff}). In this way
a direct comparison between equilibrium nonzero temperature and non--equilibrium 
nonzero voltage bias is possible: Naively, one might expect that not only the zeroth order
result (\ref{chi0ParHol}) can be described by the effective temperature (\ref{eqdefTeff}), 
but that this also holds for the logarithmic correction in (\ref{eqchi0logcorr}).
This is clearly not the case and the logarihmic corrections to $\chi_0(V)$
contain some genuine non--equilibrium physics.\cite{footnote_efftemp} Due to
numerical limitations it has unfortunately not been possible to go to larger
values of the voltage bias to study the full crossover to the analytically known
asymptotic behavior $\lim_{V\rightarrow\infty} V\chi_0(V)/4=1/4$. 

Fig.~\ref{Figchi0RTeff} contains the non--equilibrium static spin susceptibility 
plotted in the same way for various values of the asymmetry parameter~$r$. One can
see that the logarithmic corrections get even larger for increasing asymmetry.
The values of $T_{\rm eff}\,\chi_0(T_{\rm eff})$ clearly start to overshoot the
asymptotic value $1/4$ for larger asymmetries. As opposed to the equilibrium
case, the effective magnetic moment in non--equilibrium first increases upon
lowering the voltage bias. This is related to the strong energy-dependence of the
running coupling constant in non--equilibrium, see also Figs.~\ref{fig_flowup_sym} 
and \ref{fig_flowup_asym}. The behavior here is consistent with what one would
conjecture based on the analytically known $\ln^2$--corrections for a Kondo model that
cannot be derived from an underlying Anderson impurity model.
In that case
the energy--dependence of the couplings becomes even larger than here, eventually reaching 
two--channel Kondo physics behavior. Likewise, the leading logarithmic corrections to 
the static spin susceptibility are positive,\cite{Rosch03,Rosch04a,Rosch05}  which is reminiscent of our results
depicted in Fig.~\ref{Figchi0RTeff}.

The behavior of the curves for smaller values of the voltage bias in Fig.~\ref{Figchi0RTeff}
has to be interpreted with care, since the running coupling constants become larger
and therefore the flow equation calculation less reliable. Still, the fact that the
curves for different asymmetries converge for smaller voltage bias has a simple 
physical interpretation: We know that $\chi_0(V=0,T=0)=w/4T_{\rm K}$ with the 
Wilson number $w=0.413$ known exactly from the Bethe ansatz. This result is 
trivially independent of~$r$. This translates into a constant slope~$w/4$
in Fig.~\ref{Figchi0RTeff} (dashed line), which all the other curves have
to approach in the limit $V\ll T_{\rm K}$ independent of~$r$. Notice that 
our results for the non--equilibrium static spin susceptibility show clear 
indications of this behavior. 

\section{Conclusions}
This paper has served a twofold purpose. First, we have worked out
how the flow equation method can be applied to steady state non--equilibrium
impurity models. This was illustrated in detail for the Kondo model with voltage bias.
Thereby we have shown how Hamiltonian scaling ideas can be used in
non--equilibrium, and how they help to understand the physics qualitatively
and also quantitatively. A point of particular importance is the transmutation
of marginal terms in the Hamiltonian into irrelevant terms on the energy scale set by the spin
relaxation rate, which is generated by the steady state current. This is precisely 
the effect of decoherence within our Hamiltonian scaling framework. 

In addition to these methodological considerations, we have also derived a number of
new results for the non--equilibrium spin dynamics in the Kondo model
with voltage bias. In this paper we have focussed on the spin dynamics
since it contains frequency resolved information about the non--equilibrium
steady state (as opposed to the steady state current (\ref{eq_GV})). For
the equilibrium model the spin dynamics is the key to understanding Kondo physics,
and therefore the spin dynamics is a very suitable tool to illustrate the 
differences between equilibrium and non--equilibrium. In particular, we have
calculated the static spin susceptibility in non--equilibrium including its
logarithmic corrections for a wide range of voltage bias. Although we had to 
resort to a numerical solution
of the flow equations for this purpose, this is an important step since to
date an analytical calculation of this leading correction has not been
possible (at least for the experimentally most relevant case of a Kondo
model that is derived from an underlying Anderson impurity model).

Scaling ideas are one of the key concepts for understanding equilibrium
many--body systems. Likewise, 
we believe that Hamiltonian scaling ideas like the one presented here
can play an equally important role for steady state non--equilibrium problems.

\begin{acknowledgments}
We acknowledge valuable discussions with 
N.~Andrei, J.~von Delft, A.~Mitra, J.~Paaske and A.~Rosch.
This work was supported through SFB/TR~12 of the Deutsche 
Forschungsgemeinschaft (DFG), the Center for Nanoscience (CeNS)
Munich, and the German Excellence Initiative via the
Nanosystems Initiative Munich (NIM).
\end{acknowledgments}

\appendix \section{Commutators and Normal--Ordering}
In this Appendix we work out in detail some key commutators that are used
throughout our calculations.

~~\newline
\noindent
{\bf 1. Commutator ${ [\vec S \cdot \vec s^\pdag_{t't}\, , 
\,\vec S \cdot \vec s^\pdag_{u'u}]}$} \newline
~~\newline
We use the following fundamental property of the spin--1/2 algebra
\beq
S^i\,S^j=\frac{1}{4}\,\delta_{ij}+\frac{i}{2}\sum_k \epsilon_{ijk}\,S^k
\label{def_S}
\eeq
and insert it into the commutator:
\bea
\lefteqn{
[\vec S \cdot \vec s^\pdag_{t't}\, , \,\vec S \cdot \vec s^\pdag_{u'u}] }
\label{eq_app1} \\
&& \nn \\
&=& \frac{1}{4} \sum_{i,j} \Big( S^i S^j c^\dag_{t'\alpha}
  \sigma^i_{\alpha\beta} c^\pdag_{t\beta}\, 
c^\dag_{u'\mu} \sigma^j_{\mu\nu} c^\pdag_{u\nu} \nn \\
&& \qquad - S^j S^i c^\dag_{u'\mu}
  \sigma^j_{\mu\nu} c^\pdag_{u\nu}\, 
c^\dag_{t'\alpha} \sigma^i_{\alpha\beta} c^\pdag_{t\beta}  \Big) \nn \\
&=&\frac{i}{8} \sum \epsilon^\pdag_{ijk}  \sigma^i_{\alpha\beta}
 \sigma^j_{\mu\nu} S^k \left( 
c^\dag_{t'\alpha} c^\pdag_{t\beta} c^\dag_{u'\mu} c^\pdag_{u\nu}
+ c^\dag_{u'\mu} c^\pdag_{u\nu} c^\dag_{t'\alpha} c^\pdag_{t\beta} \right) \nn \\
&&+\frac{1}{16} \sum_i ( \sigma^i_{\alpha\beta} \sigma^i_{\mu\nu} )
\,[c^\dag_{t'\alpha} c^\pdag_{t\beta}, c^\dag_{u'\mu} c^\pdag_{u\nu}] \nn \\
&=&\frac{i}{8} \sum \epsilon^\pdag_{ijk}  \sigma^i_{\alpha\beta}
 \sigma^j_{\mu\nu} S^k \left( 
c^\dag_{t'\alpha} c^\pdag_{t\beta} c^\dag_{u'\mu} c^\pdag_{u\nu}
+ c^\dag_{u'\mu} c^\pdag_{u\nu} c^\dag_{t'\alpha} c^\pdag_{t\beta} \right) \nn
\\
&&+\frac{3}{16}\sum_\alpha \left(\delta^\pdag_{tu'}\,c^\dag_{t'\alpha} c^\pdag_{u\alpha} 
-\delta^\pdag_{t'u}\,c^\dag_{u'\alpha} c^\pdag_{t\alpha} \right) \nn 
\eea
Next we need to normal--order the fermion terms. We introduce the following
expectation values with respect to the non--interacting ground state
\bea
n^+(u) &\stackrel{\rm def}{=}& \langle c^\dag_{u\alpha} c^\pdag_{u\alpha}
\rangle = n(u)
\label{def_n} \\
n^-(u) &\stackrel{\rm def}{=}& \langle c^\pdag_{u\alpha} c^\dag_{u\alpha} \rangle \nn
\eea
which are related by $n^+(u)=1-n^-(u)$ because of 
$\{c^\dag_u,c^\pdag_{u'}\}=\delta_{uu'}$. No summation over $\alpha$ is
implied in (\ref{def_n}), and the expectation values are obviously independent
of $\alpha$ due to spin symmetry (no magnetic field).  
Then
\bea
:c^\dag_{u'\alpha} c^\pdag_{u\beta}: &\stackrel{\rm def}{=}&
c^\dag_{u'\alpha} c^\pdag_{u\beta} -\delta^\pdag_{u'u}
\delta^\pdag_{\alpha\beta} \,n^+(u) \\
:c^\pdag_{u\beta} c^\dag_{u'\alpha}: &\stackrel{\rm def}{=}&
c^\pdag_{u\beta} c^\dag_{u'\alpha} -\delta^\pdag_{u'u}
\delta^\pdag_{\alpha\beta} \,n^-(u) \nn
\eea
and $:c^\dag_{u'\alpha} c^\pdag_{u\beta}:=-:c^\pdag_{u\beta} c^\dag_{u'\alpha}:$.
For normal--ordering of higher--order fermion terms we follow the prescription
of Wegner~\cite{Wegner94} that not only subtracts the ground state expectation
values, but also expectation values with respect to lower--order excited
states:
\bea
:c^\dag_{t'\alpha} c^\pdag_{t\beta} c^\dag_{u'\mu} c^\pdag_{u\nu}:
&\stackrel{\rm def}{=}&
:c^\dag_{t'\alpha} c^\pdag_{t\beta}: \;  :c^\dag_{u'\mu} c^\pdag_{u\nu}:
\label{nn4} \\
&&-:c^\dag_{t'\alpha} c^\pdag_{u\nu}:\, \delta_{tu'} \delta_{\beta\mu} n^-(t)
\nn \\
&&+:c^\dag_{u'\mu} c^\pdag_{t\beta}: \, \delta_{t'u} \delta_{\alpha\nu} n^+(u)
\nn \\
&&- \delta_{t'u}\delta_{tu'} \delta_{\alpha\nu}\delta_{\beta\mu} n^-(t)  n^+(u)
\nn
\eea
Inserting everything into (\ref{eq_app1}) yields after some straightforward
algebra
\bea
\lefteqn{
[\vec S \cdot \vec s^\pdag_{t't}\, , \,\vec S \cdot \vec s^\pdag_{u'u}] }
\label{app_comm1} \\
&=&i\, :\vec S\cdot (\vec s^\pdag_{t't} \times \vec s^\pdag_{u'u}): 
\nn \\
&&+\vec S\cdot \vec s^\pdag_{t'u} \,\delta^\pdag_{tu'}
(n^+(t)-1/2) \nn \\
&&- \vec S\cdot \vec s^\pdag_{u't} \,\delta^\pdag_{t'u}
(n^+(u)-1/2) \nn \\
&&+\frac{3}{16}\sum_\alpha \left(
\delta^\pdag_{tu'}\, :c^\dag_{t'\alpha} c^\pdag_{u\alpha}:
-\delta^\pdag_{t'u}\, :c^\dag_{u'\alpha} c^\pdag_{t\alpha}: \right) \nn \\
&&+\frac{3}{8}\, \delta^\pdag_{tu'} \delta^\pdag_{t'u}
\,(n^+(u)-n^+(t)) \nn \ .
\eea
Normal--ordering in the first term on the rhs of this equation acts
on the fermions only
\beq
 :\vec S\cdot (\vec s^\pdag_{t't} \times \vec s^\pdag_{u'u}): 
= \vec S\,\cdot\; :(\vec s^\pdag_{t't} \times \vec s^\pdag_{u'u}): \ .
\eeq

~~\newline
\noindent
{\bf 2. Commutator ${ [:\vec S\cdot (\vec s^\pdag_{1'1} \times 
\vec s^\pdag_{2'2}):\, , \, \vec S \cdot \vec s^\pdag_{4'4}] }$} \newline 
~~\newline
For notational clarity we use labels $1',1,2',2,4',4$ instead of
$v',v,w',w,u',u$ in this section. For the purposes of this paper
we also only need to identify terms with the structure 
$\vec S \cdot \vec s^\pdag_{t't}$ in the above normal--ordered
commutator. This will simplify our calculation considerably. 

Using (\ref{def_S}) it is easy to show
\bea
\lefteqn{
[:\vec S\cdot (\vec s^\pdag_{1'1} \times 
\vec s^\pdag_{2'2}):\, , \, \vec S \cdot \vec s^\pdag_{4'4}] } 
\label{eq_app2} \\
&=&\frac{1}{4}\sum_{a=1}^3  
[:(\vec s^\pdag_{1'1} \times 
\vec s^\pdag_{2'2})^a :\, , \,  s^a_{4'4}] \nn \\
&&+\frac{i}{2}\sum_{a,b,c=1}^3 \eps_{abc} S^c
\{:(\vec s^\pdag_{1'1} \times 
\vec s^\pdag_{2'2})^a :\, , \,  s^b_{4'4} \} \nn
\eea
The first term does not contain the impurity spin and can therefore not
contribute to the terms that we need to extract from the commutator.
We can focus on the anticommutator in the second term
\bea
\lefteqn{
\sum_{a,b=1}^3 \eps_{abc} 
\{:(\vec s^\pdag_{1'1} \times 
\vec s^\pdag_{2'2})^a :\, , \,  s^b_{4'4} \} } 
\label{anticomm} \\
&=&\sum_{i=1}^3 \{:s^i_{1'1}  
s^c_{2'2} :\, , \,  s^i_{4'4} \}
-\sum_{i=1}^3 \{:s^c_{1'1}  
s^i_{2'2} :\, , \,  s^i_{4'4} \} \nn
\eea
In order to yield terms with the structure $\vec S \cdot \vec s^\pdag_{t't}$
we need to extract the terms with two contractions (i.e., two $n^\pm$--terms)
in (\ref{anticomm}). Similar to (\ref{nn4}) one shows
\bea
\lefteqn{
:c^\dag_{1'} c^\pdag_1 c^\dag_{2'} c^\pdag_2: \: :c^\dag_{4'} c^\pdag_4: } \\
&=&{\rm :(0-contraction):~+~:(1-contraction):} \nn \\
&&+\delta_{1'4}\delta_{14'}\, n^+(1') n^-(1)\, :c^\dag_{2'} c^\pdag_2: \nn \\
&&+\delta_{2'4}\delta_{24'}\, n^+(2') n^-(2)\, :c^\dag_{1'} c^\pdag_1: \nn \\
&&-\delta_{1'4}\delta_{24'}\, n^+(1') n^-(2)\, :c^\dag_{2'} c^\pdag_1: \nn \\
&&-\delta_{2'4}\delta_{14'}\, n^+(2') n^-(1)\, :c^\dag_{1'} c^\pdag_2: \nn
\eea
From this expression it is straightforward to show
\bea
\lefteqn{
\sum_{i=1}^3 :s^i_{1'1}  
s^c_{2'2}:\,  s^i_{4'4} } \nn \\ 
&=&
{\rm :(0-contraction):~+~:(1-contraction):} \nn \\
&&+\delta_{1'4}\delta_{14'}\, n^+(1') n^-(1)\,\frac{3}{2} s^c_{2'2} \nn \\
&&+\delta_{2'4}\delta_{24'}\, n^+(2') n^-(2)\,\frac{1}{2} s^c_{1'1} \nn \\
&&-\delta_{1'4}\delta_{24'}\, n^+(1') n^-(2)\,\frac{3}{4} s^c_{2'1} \nn \\
&&-\delta_{2'4}\delta_{14'}\, n^+(2') n^-(1)\,\frac{3}{4} s^c_{1'2} \nn
\eea
Combining all the like terms in (\ref{anticomm}) yields
\bea
\lefteqn{
\sum_{a,b=1}^3 \eps_{abc} 
\{:(\vec s^\pdag_{1'1} \times 
\vec s^\pdag_{2'2})^a :\, , \,  s^b_{4'4} \} } \\
&=&{\rm :(0-contraction):~+~:(1-contraction):} \nn \\
&&+\delta_{1'4}\delta_{14'}\, (n^+(1') n^-(1)+n^+(1) n^-(1')) \, s^c_{2'2} \nn \\
&&-\delta_{2'4}\delta_{24'}\, (n^+(2') n^-(2)+n^+(2) n^-(2')) \, s^c_{1'1} \nn
\eea
This gives our desired result by inserting into (\ref{eq_app2})
\bea
\lefteqn{
[:\vec S\cdot (\vec s^\pdag_{1'1} \times 
\vec s^\pdag_{2'2}):\, , \, \vec S \cdot \vec s^\pdag_{4'4}] } 
\label{app_comm2} \\
&=&\frac{i}{2}\,\delta_{1'4}\delta_{14'}\, 
(n^+(1') n^-(1)+n^+(1) n^-(1'))\, \vec S\cdot \vec s^\pdag_{2'2} \nn \\
&&-\frac{i}{2}\,\delta_{2'4}\delta_{24'}\, 
(n^+(2') n^-(2)+n^+(2) n^-(2')) \, \vec S\cdot \vec s^\pdag_{1'1}
\nn \\
&&+{\rm normal-ordered~terms~with~different~structure} \nn
\eea

\section{Numerical Solution}
\label{app_numerics}
This Appendix contains some details of the numerical solution of
the full set of flow equations (\ref{neq_J2loopa}) and (\ref{neq_K2loopa}).
The latter are solved using a standard Runge--Kutta algorithm\cite{NR}.

One numerical issue arises from the fact that the $K$--couplings
in (\ref{neq_K2loop}) depend on four momenta which leads to a
large number of differential equations to keep track of. While 
this is no fundamental problem, it would limit the possible resolution 
on a standard workstation considerably. However, one can use the
following approximation which reduces the complexity to quadratic
in the number of momenta and provides an excellent approximation
to the full system. 

The formal solution of (\ref{neq_K2loopa}) is given by
\bea
K^\pdag_{p'p,q'q} & = & -(\eps_{q'}-\eps_{q}) e^{-B(\eps_{p'}-\eps_{p}+\eps_{q'}-\eps_{q})^2} \\
& \ & \times \int\limits_0^B d\tilde{B}\; e^{\tilde{B}(\eps_{p'}-\eps_{p}+\eps_{q'}-\eps_{q})^2} J^\pdag_{p'p}(\tilde{B})J^\pdag_{q'q}(\tilde{B}).\nn
\eea
Using this result the flow equations for the running coupling (\ref{neq_J2loopa}) in diagonal parametrization
\beq\label{diag_param_app}
J^\pdag_{p'p} = u_{\overline{p'p}} e^{-B(\eps_{p'}-\eps_{p})^2}
\eeq
are easily rewritten to the form
\bea
\frac{du^\pdag_{p}}{dB} & = & \sum_{q} (\eps_{p}-\eps_{q})u_{\overline{pq}}^2e^{-2B(\eps_{p}-\eps_{q})^2}(2n_f(q)-1) \nn\\
& \ & -\sum_{q',q} (n_f(q')+n_f(q)-2n_f(q')n_f(q))   \nn\\
& \ & \times (\eps_{q'}-\eps_{q})^2 e^{-2B(\eps_{q'}-\eps_{q})^2} u^\pdag_{\overline{q'q}} L^\pdag_{p,\overline{q'q}}\ ,
\eea
where
\beq
L^\pdag_{p',p} = \int\limits_0^B d\tilde{B}\; u_{p'}(\tilde{B})u_{p}(\tilde{B})\ .
\eeq
In the remaining flow equations, for example the transformation of the spin operator, the running coupling $J_{p'p}(B)$
is simply replaced by the rhs of (\ref{diag_param_app}).
The equations above can be easily generalized to the case of general parameters (\ref{neq_J2loop}) and (\ref{neq_K2loop}). 

Using diagonal parametrization in the numerical solution effectively
reduces the number of differential equations to quadratic in the number of
momenta $N$. The runtime is proportional to $N^3$ instead of $N^4$ in the full set.
Additionally the stiffness of the differential equations is reduced by removing the exponential decay from the
flow of the running coupling. On a standard workstation $N$ can be choosen of ${\cal O}(1000)$.


\begin{thebibliography}{11}

\bibitem{Rosch03}
A. Rosch, J. Paaske, and P. W{\"o}lfle, Phys. Rev. Lett. {\bf 90}, 076804 (2003).
\bibitem{Rosch05}
A. Rosch, J. Paaske, J. Kroha, and P. W\"olfle, J. Phys. Soc. Jpn. {\bf 74}, 118 (2005).
\bibitem{Schoeller00}
H. Schoeller, Lect. Notes Phys. {\bf 544}, 137 (2000).
\bibitem{Schoeller07}
S. G. Jakobs, V. Meden, and H. Schoeller, Phys. Rev. Lett. {\bf 99}, 150603 (2007).
\bibitem{Mitra}
A. Mitra and A. J. Millis, Phys. Rev. B {\bf76}, 085342 (2007).
\bibitem{Millis}
D. Segal, D. R. Reichman, and A. J. Millis, Phys. Rev. B {\bf 76}, 195316 (2007).
\bibitem{Kehrein04}
S. Kehrein, Phys. Rev. Lett. {\bf 95}, 056602 (2005).
\bibitem{HacklKehrein}
A. Hackl and S. Kehrein, Phys. Rev. B {\bf 78}, 092303 (2008); arXiv:0809.3524
\bibitem{GoldhaberGordon98}
D. Goldhaber-Gordon, {\it et al.}, Nature {\bf 391}, 156 (1998).
\bibitem{Cronenwett98}
S. M. Cronenwett, T. H. Oosterkamp, and L. P. Kouwenhoven,
Science {\bf 281}, 540 (1998).
\bibitem{Schmid98}
J. Schmid, J. Weis, K. Eberl, and K. von Klitzing, Physica B{\bf 258},
182 (1998).
\bibitem{Wiel00}
W. G. van der Wiel {\it et al.}, Science {\bf 289}, 2105 (2000).
\bibitem{Glazman88}
L. Glazman and M. Raikh, JETP Letters {\bf 47}, 452 (1988).
\bibitem{Ng88}
T. Ng and P. A. Lett, Phys. Rev. Lett. {\bf 61}, 1768 (1988).
\bibitem{Glazman99} A. Kaminski, Yu. V. Nazarov and L. I. Glazman,
Phys. Rev. Lett. {\bf 83}, 384 (1999); Phys. Rev. B {\bf 62}, 8154 (2000).
\bibitem{Rosch01}
A. Rosch, J. Kroha, and P. W{\"o}lfle, Phys. Rev. Lett. {\bf 87}, 156802 (2001).
\bibitem{Rosch04a}
J. Paaske, A. Rosch, and P. W{\"o}lfle, Phys. Rev. B {\bf 69}, 155330 (2004).
\bibitem{Rosch04b}
J. Paaske, A. Rosch, J. Kroha, and P. W{\"o}lfle,
Phys. Rev. B {\bf 70}, 155301 (2004).
\bibitem{Anders08}
F. B. Anders, Phys. Rev. Lett. 101, 066804 (2008).
\bibitem{Schmitteckert}
E. Boulat, H. Saleur, and P. Schmitteckert, Phys. Rev. Lett. {\bf 101}, 140601 (2008).
\bibitem{Andrei}
P. Mehta and N. Andrei, Phys. Rev. Lett. {\bf 96}, 216802 (2006); 
P. Mehta, Sung po Chao, and N. Andrei, arXiv:cond-mat/0703426 
\bibitem{MoeckelKehrein}
M. Moeckel and S. Kehrein, Phys. Rev. Lett. {\bf 100}, 175702 (2008).
\bibitem{Parcollet02}
O. Parcollet and C. Hooley, Phys. Rev. B {\bf 66}, 085315 (2002).
\bibitem{Hewson}
For an overview see, e.g., A. C. Hewson, 
{\it The Kondo Problem to Heavy Fermions} (Cambridge Univ.\ Press, 1993).
\bibitem{Wilson_NRG}
K. G. Wilson, Rev. Mod. Phys. {\bf 47}, 773 (1975).
\bibitem{Wegner94} 
F. Wegner, Ann. Physik (Leipzig) {\bf 506}, 77 (1994).
\bibitem{Wilson93}
S. G{\l}azek and K. G. Wilson, Phys. Rev. D {\bf 48}, 5863 (1993); 
{\it ibid.} {\bf 49}, 4214 (1994).
\bibitem{Kehreinbook}
S. Kehrein, {\it The Flow Equation Approach to Many-Particle Systems} (Springer, Berlin, 2006).
\bibitem{Anderson70}
P. W. Anderson, J. Phys. C {\bf 3}, 2436 (1970).
\bibitem{Kehrein96}
S. Kehrein and A. Mielke, Ann. Physik (Leipzig) {\bf 6}, 90 (1997).
\bibitem{Kleff}
S. Kleff, S. Kehrein, and J.~von Delft, 
Phys. Rev. B {\bf 70}, 014516 (2004).
\bibitem{Wegner_Hubbard}
I. Grote, E. K{\"o}rding, and F. Wegner,
J. Low Temp. Phys. {\bf 126}, 1385 (2002);
V. Hankevych, I. Grote, and F. Wegner,
Phys. Rev. B {\bf 66}, 094516 (2002).
\bibitem{Uhrig}
C. Knetter, K.~P.~Schmidt, M. Gr{\"u}ninger, and G.~S.~Uhrig,
Phys. Rev. Lett. {\bf 87}, 167204 (2001);
C. Knetter, K.~P.~Schmidt, and G.~S.~Uhrig, 
Eur. Phys. J. B {\bf 36}, 525 (2004).
\bibitem{Kehrein_SG} 
S. Kehrein, Phys. Rev. Lett. {\bf 83}, 4914 (1999);
Nucl. Phys. B[FS] {\bf 592}, 512 (2001).
\bibitem{Kehrein_KM}
W. Hofstetter and S. Kehrein,
Phys. Rev. B {\bf 63}, 140402(R) (2001).
\bibitem{Garst}
M. Garst {\it et al.}, Phys. Rev. B {\bf 69}, 214413 (2004).
\bibitem{LobaskinKehrein}
D. Lobaskin and S. Kehrein, Phys. Rev. B {\bf  71}, 193303 (2005).
\bibitem{Coleman01}
P. Coleman, C. Hooley, and P. Parcollet, Phys. Rev. Lett. {\bf 86},
4088 (2001).
\bibitem{Note_expparameters}
Notice that it is possible to have flow equation expansion parameters
different (i.e., more well--behaved) from the running coupling constants,
which is the reason why flow equations can yield a controlled expansion
even in certain strong--coupling problems~\cite{Kehrein_SG,Kehrein_KM}. 
\bibitem{footnote_us} 
The differential equations for $u_t, u_l, u_r$ correspond to Eqs.~(5) and (6)
in Ref.~\protect\onlinecite{Kehrein04}.
\bibitem{footnote_finiteT}
One can work this out easily from (\ref{eq_J2loop}) 
and (\ref{eq_K2loop}) with finite temperature Fermi functions~$n(u)$.

\bibitem{BA_thermo}
V. M. Filyov, A. M. Tzvelik, and P. B. Wiegmann, Physics Letters A {\bf 81}, 175 (1981).
\bibitem{BA_scales}
N. Andrei and J. H. Lowenstein, Phys. Rev. Lett. {\bf 46}, 356 (1981).
\bibitem{NR}
See e.g. W. H. Press {\it et al.}, {\it Numerical Recipes, Third Edition} (Cambridge Univ. Press, 2007).
\bibitem{FlucDiss}
H. B. Callen and T. A. Welton, Phys. Rev. {\bf 83}, 34 (1951).
\bibitem{footnote_efftemp}
Another way of saying this is to argue that the effective temperature
acquires logarithmic corrections in non--equilibrium. However, one would
need to check other observables to verify whether this notion is useful.

\end{thebibliography}
\end{document}